\documentclass{article}

\usepackage{latexsym}

\newcommand{\eqnum}[1]{\setcounter{equation}{#1}} 

\begin{document}
\begin{titlepage}
\baselineskip .15in
\begin{flushright}
WU-AP/132/01
\end{flushright}

\begin{center}
{\bf

\vskip 1cm
{\Large Asymptotic symmetries on Killing horizons}

}\vskip .6in

{\sc  Jun-ichirou Koga}${}^{\dagger}$\\[1em]

{\em Advanced Research Institute for Science and Engineering, 

Waseda University, Shinjuku-ku, Tokyo 169-8555, Japan} 

\end{center}
\vfill
\begin{abstract} 
We investigate asymptotic symmetries regularly defined 
on spherically symmetric Killing horizons in the 
Einstein theory with or without the cosmological constant. 
Those asymptotic symmetries are described by 
asymptotic Killing vectors, along which the Lie derivatives of 
perturbed metrics vanish on a Killing horizon. 
We derive the general form of asymptotic Killing vectors 
and find that the group of the asymptotic symmetries consists of 
rigid $O(3)$ rotations of a horizon two-sphere and 
supertranslations along the null direction on the horizon,  
which depend arbitrarily 
on the null coordinate as well as the angular coordinates. 
By introducing the notion of asymptotic Killing horizons, 
we also show that local properties of Killing horizons 
are preserved under not only diffeomorphisms but also non-trivial 
transformations generated by the asymptotic symmetry group. 
Although the asymptotic symmetry group contains the 
$\mathit{Diff}(S^1)$ subgroup, which results from 
the supertranslations dependent only on the null coordinate, 
it is shown that the Poisson bracket algebra of 
the conserved charges conjugate to asymptotic Killing vectors 
does not acquire non-trivial central charges. 
Finally, by considering extended symmetries, we discuss that 
unnatural reduction of the symmetry group is necessary 
in order to obtain the Virasoro algebra with non-trivial central charges, 
which will not be justified when we respect 
the spherical symmetry of Killing horizons. 
\end{abstract}

\vskip 1cm
\begin{center}
July, 2001
\end{center}
\vfill
\baselineskip .2in
$\dagger$~electronic mail : koga@gravity.phys.waseda.ac.jp
\end{titlepage}

\normalsize

\baselineskip 15pt
\section{Introduction} 
\label{sec:Introduction} 
\eqnum{0} 

The recent progress in string theory, 
such as the techniques of D-branes\cite{Dbrane} and 
the AdS/CFT correspondence\cite{AdSCFT}, has laid 
the foundations of the successful methods 
to derive the Bekenstein--Hawking formula for 
the entropy of a black hole in microscopic manners. 
However, those methods 
rely on the particular features to string theory, and then 
we cannot apply them to generic black holes. 
In addition, it is not clearly discussed in string theory 
where the microscopic states responsible 
for the black hole entropy live, 
while it has been expected that the entropy of a black hole will 
be associated with local or quasi-local properties of 
the black hole horizon. 
These two issues will be closely related, because the universal feature 
in black hole spacetimes is presence of black hole horizons 
and its role is obscure in those methods of string theory. 

In this respect, the quasi-local analyses
\cite{TrappingHorizon,IsolatedHorizon,WeaklyIsolated} of black hole 
horizons will provide helpful insights into the black hole entropy. 
Indeed, in the context of isolated horizons\cite{IsolatedHorizon}, 
it has been shown\cite{LQGentropy} for a wide class of black holes 
that the entropy calculated by the method based on loop quantum gravity 
is proportional to the area of the black hole. 
Since isolated horizons are characterized by the (quasi-)local 
properties of black holes, microscopic physics is associated with 
the local properties of black hole horizons by this method. 
Unfortunately, however, loop quantum gravity contains 
an undetermined parameter, called Immirzi parameter, 
and so does the expression of the black hole entropy. 

Thus, we still need to develop microscopic theories of black holes 
in order to establish a universal method applicable to generic 
black holes, which is required to correctly reproduce 
the Bekenstein--Hawking formula in a microscopic manner 
and be closely related to local properties of black hole horizons. 
One possibility will be to apply and extend the successful method, 
by which we can derive microscopically the entropy of 
the B.T.Z. black hole\cite{Strominger}. 
This method is based on the results of Ref.\cite{BrownHenneaux}, 
where it has been shown  
that deformations of the surface at the spatial infinity 
are generated by the group of asymptotic symmetries and 
are described by the 2-d conformal field theory. 
The Lie bracket algebra of those asymptotic symmetries is then 
given by two copies of the $\mathit{Diff}(S^1)$ algebra, 
\begin{equation} 
\bigl[ \zeta_n \, , \, \zeta_m \bigr]^{\mu} = i \: (m - n) \, 
\zeta^{\mu}_{n+m} \; ,  
\label{eqn:DiffS1BTZ} 
\end{equation} 
and the Poisson bracket algebra (in the reduced phase space) 
is described by two copies of the Virasoro algebra 
(by replacing the commutators with the Poisson brackets), 
\begin{equation} 
i \, \Bigl\{ \mbox{\large ${\cal H}$}[g; \zeta_n] \, , \, 
\mbox{\large ${\cal H}$}[g; \zeta_m] \Bigr\} = (n - m) \, 
\mbox{\large ${\cal H}$}[g; \zeta_{n+m}] 
+ (n^3 - n) \, \frac{c}{12} \, \delta_{n+m, 0} \; , 
\label{eqn:Virasoro}  
\end{equation} 
with the central charge $c = 3 \: l / 2 \, G$, 
where the vectors $\zeta^{\mu}_n$ generate the asymptotic symmetries, 
$\mbox{\large ${\cal H}$}[g; \zeta_n]$ are the conserved charges 
conjugate to $\zeta^{\mu}_n$ in a spacetime $g_{\mu \nu}$, 
and $l$ is related to the cosmological 
constant $\Lambda$ by $l^2 = - 1 / \Lambda$. 
Then, by applying Cardy's formula, 
the number of the microscopic states in 
the corresponding quantum theory is expressed 
in terms of the central charge $c$, 
which correctly reproduces the Bekenstein--Hawking formula 
for the B.T.Z. black hole. 
This method is applicable also to higher-dimensional stringy 
black holes, 
and hence it provides a reliable basis of the AdS/CFT correspondence. 

Although the above method is based on the asymptotic symmetries 
at infinity in the 3-d anti-de Sitter spacetime, 
it is a tempting idea that 
we consider asymptotic symmetries on generic black hole horizons.  
When we recall that an event horizon is 
a conformally invariant structure, 
we expect that there may exist asymptotic symmetries 
on black hole horizons, 
and those are described by the conformal field theory. 
It may be misleading to call such symmetries on black hole horizons 
``asymptotic'' symmetries, since a black hole horizon generically locates 
at a finite physical distance from the exterior region. 
However, here we call them asymptotic symmetries, if any, 
because those are expected to shear some features 
with usual, genuine asymptotic symmetries. 

As to usual asymptotic symmetries, 
those associated with the B.M.S. group\cite{BMS} and 
the Spi group\cite{Spi} are also well-known, which are defined respectively at the null and spacelike infinity 
in a 4-d asymptotically flat spacetime. 
Those asymptotic symmetries are closely related to 
the asymptotic behavior of the spacetime geometry. 
In addition, a remarkable feature common to 
those asymptotic symmetries is that they are generated 
by the larger groups 
than the isometry groups in the corresponding 
maximally symmetric spacetimes. 
Thus, by a method based on asymptotic symmetries, 
it will be possible to associate 
a large number of degrees of freedom, which may describe 
microscopic physics responsible for the entropy of a black hole, 
with the local geometry near the black hole horizon. 
In fact, a large number of degrees of freedom will be necessary for 
the entropy of a sufficiently large, semi-classical black hole. 
We then expect that analyses of asymptotic symmetries 
on black hole horizons may provide a clue to establish 
a universal method to derive the Bekenstein--Hawking formula. 

Carlip has analyzed asymptotic symmetries on black hole horizons 
in two papers\cite{Carlip1,Carlip2}, 
and argued that the Poisson bracket algebra associated with those 
asymptotic symmetries gives the Virasoro algebra, 
which reproduces the Bekenstein--Hawking formula. 
However, both of these papers have been criticized 
for the errors in the formulations. The functional derivatives of 
conserved charges are not well-defined\cite{ParkHo} 
in the first paper (see also Ref.\cite{CarpipRep}), 
and the generating vectors of asymptotic symmetries do not 
satisfy the closure condition of the Lie bracket algebra\cite{DreyerEtAl} 
in the second paper. 
Thus, the conceptually plausible idea of 
Carlip to derive the Bekenstein--Hawking formula 
from asymptotic symmetries on black hole horizons 
has not been successfully realized so far. 
However, these errors seem related to the singular behavior of 
the asymptotic forms of the relevant quantities. 
We should take care when we analyze quantities defined 
on black hole horizons. 
We also need to pay attention to conserved charges 
on a Cauchy surface, 
since the Cauchy surfaces defined by constant Killing time are 
generically degenerate on the bifurcation surface of a Killing horizon. 

Then, in this paper, we consider asymptotic symmetries regularly 
defined on black hole horizons, as well as cosmological horizons, 
which are known to shear thermodynamical properties 
with black hole horizons\cite{GibbonsHawking}.   
We also consider how the asymptotic symmetries 
are related to local properties of those horizons, 
and whether they are responsible for the Virasoro algebra. 
To analyze those issues, we focus on Killing horizons in 
4-d spacetimes described by the Einstein theory with or without 
the cosmological constant. 
Although event horizons in dynamically evolving spacetimes are not 
Killing horizons generally, 
the properties of Killing horizons are expressed 
in the local forms, which suits to our present purpose. 
Furthermore, we will be concerned only with 
spherically symmetric spacetimes. 
This is because presence of angular momentum in four dimensions 
may obscure generality of the discussion. 
The direction specified by the angular momentum 
together with the Killing time defines a preferred 2-d surface, 
on which we may happen to find the 2-d conformal group. 
However, this method does not work 
in general, since such a preferred 2-d surface does not exist 
in the spherically symmetric case. We need a formulation which  
works {\it even} in the spherically symmetric case.

Thus, in Section \ref{sec:AsymptoticSymmetries}, 
we begin with local geometric features  
near spherically symmetric Killing horizons, along with the examples 
of the black hole horizon in the Schwarzschild spacetime and 
the cosmological horizon in the de Sitter spacetime. 
We then consider the asymptotic symmetries on spherically 
symmetric Killing horizons, which are generated 
by asymptotic Killing vectors. 
By deriving the general form of asymptotic Killing vectors, 
we analyze features of the asymptotic symmetries 
on Killing horizons. 
The explicit forms of asymptotic Killing vectors and related 
results in this paper are presented only in the case of the future cosmological horizon in the de Sitter spacetime. 
This is because we will discuss the structure of the group of the 
asymptotic symmetries by comparing with the isometry group 
in the maximally symmetric spacetime, 
as the B.M.S. group is compared with the Poincar\'e group. 
However, as we will describe, all the analyses in this paper, 
except for the comparison with the de Sitter group, 
are readily applicable to black hole horizons. 

We analyze in Section \ref{sec:AsymptoticKillingHorizon} 
how the asymptotic symmetries on Killing horizons 
are related to local properties of Killing horizons. 
We first introduce the notion of asymptotic Killing horizons, 
which shear local properties with exact Killing horizons, 
and then analyze behavior of asymptotic Killing horizons 
under transformations generated by the asymptotic symmetry group. 
While it will be trivial that any covariant properties are 
preserved under diffeomorphisms, we investigate whether asymptotic 
Killing horizons remain unaffected also under 
non-trivial transformations 
associated with the asymptotic symmetries. 

In Section \ref{sec:PoissonBracket}, we analyze the Poisson bracket 
algebra of the conserved charges conjugate to asymptotic Killing 
vectors, by using the covariant phase space formalism developed by 
Wald and his collaborators\cite{CovariantPhase,WaldZoupas}. 
Unfortunately, however, it will turn out that the Poisson bracket algebra 
does not acquire non-trivial central charges. 
It implies that 
the Poisson bracket algebra cannot contain 
the Virasoro algebra Eq.(\ref{eqn:Virasoro}) 
with non-trivial central charges,  
and hence the entropy of a generic black hole cannot be derived 
from the asymptotic symmetries on Killing horizons
by the same method as that of the B.T.Z. black hole. 
Since it is suspected that we may obtain a central charge 
if we consider a larger symmetry group, 
we then analyze extended symmetries and discuss whether and how 
we can obtain the Virasoro algebra with non-trivial central charges. 

Finally, we summarize this paper and discuss related issues in 
Section \ref{sec:Discussion}. 
The appendices are devoted to the derivation of 
asymptotic Killing vectors and the calculation of a central charge, 
as well as a list of the Killing vectors in the de Sitter spacetime. 
In what follows, we use the units $c = G = 1$.

\section{Asymptotic symmetries on Killing horizons} 
\label{sec:AsymptoticSymmetries} 
\eqnum{0}

\subsection{Local geometry near Killing horizons}

Here, we briefly review local geometric features near 
Killing horizons in static and spherically symmetric 
4-d spacetimes with the metric ${}^{0}\!g_{\mu \nu}$, 
which are referred to as 
the background spacetimes in what follows. 
We will work in a regular coordinate system 
to analyze asymptotic symmetries on Killing horizons in 
a regular manner. 

It is well-known that the metric of the Schwarzschild spacetime 
in the Kruskal coordinate system is given by 
\begin{equation} 
ds^2 = {}^{0}\!g_{\mu \nu} \; dx^{\mu} \: dx^{\nu} 
= - \, \frac{2 M}{r} \exp \! \left[ - \frac{r}{2 M} \right] du \: dv 
+ r^2 \: d \Omega^2 \; ,  
\label{eqn:MetricSchKruskal} 
\end{equation} 
where $M$ denotes the mass of the black hole, 
the Kruskal coordinates are defined by 
\begin{equation} 
u \equiv 
- \; 4 \: M \exp \! \left[ - \frac{U}{4 M} \right] 
\; , 
~~~~~~ 
v \equiv 
4 \: M \exp \! \left[ \frac{V}{4 M} \right] \; , 
\end{equation} 
and the null coordinates $U \equiv t - r^{*}$ and $V \equiv t + r^{*}$ are 
defined in terms of the Killing time coordinate $t$ and 
the tortoise coordinate $r^{*}$. 
Then, the future and past black hole event horizons are located at 
$u = 0$ and $v = 0$, respectively. 

Similarly, the metric of the de Sitter spacetime is written as 
\begin{equation}
ds^2 = {}^{0}\!g_{\mu \nu} \; dx^{\mu} \: dx^{\nu} 
= - \, \frac{4 \: \ell^4}{\left( u \: v - \ell^2 \right)^2} \; du \, dv \; 
+ \; \ell^2 \: 
\frac{\left( u \: v + \ell^2 \right)^2}{\left( u \: v - \ell^2 \right)^2} 
\; d \Omega^2 \; ,  
\label{eqn:DSKrsMetric}
\end{equation} 
by introducing the Kruskal coordinates 
\begin{equation}
u \equiv 
\ell \, \exp \! \left[ \frac{U}{\ell} \right] \; , 
~~~~~~ 
v \equiv 
- \, \ell \, \exp \! \left[ - \frac{V}{\ell} \right] \; ,  
\label{eqn:Krs} 
\end{equation}
where $U \equiv t - r^{*}$ and $V \equiv t + r^{*}$ are defined in the 
same way as above, and $\ell$ is related to the 
cosmological constant by $\ell^2 = 3 / \Lambda$. 
The future and past cosmological event horizons locate at $v = 0$ and 
$u = 0$, respectively. 

Near the future cosmological horizon ($v = 0$), 
the metric of the de Sitter spacetime Eq.(\ref{eqn:DSKrsMetric}) is 
written as 
\begin{equation}
g_{u v} = - 2 + {\cal O}(v) \; , 
~
g_{\theta \theta} = r_{\scriptscriptstyle H}^2  
+ {\cal O}(v) \; , 
~ 
g_{\phi \phi} = r_{\scriptscriptstyle H}^2 \sin^2 \! \theta 
+ {\cal O}(v) \; , 
~ 
g_{\mu \nu} = {\cal O}(v) \; 
\mbox{ for other components} \; ,   
\label{eqn:BackMetricAsym} 
\end{equation} 
where $r_{\scriptscriptstyle H} = \ell$ is the radius of 
the cosmological horizon. The Schwarzschild metric 
Eq.(\ref{eqn:MetricSchKruskal}) near the future black hole horizon 
($u = 0$) is also recast into the form of Eq.(\ref{eqn:BackMetricAsym}) 
with $r_{\scriptscriptstyle H} = 2 M$, 
by interchanging the roles of $u$ and $v$,  
and rescaling those coordinates appropriately. 
The asymptotic form of the metric near the Killing horizon 
in the static and spherically symmetric background spacetime 
will be generically  given by Eq.(\ref{eqn:BackMetricAsym}), 
while we have not proved it 
explicitly (see Ref.\cite{RaczWald} for the related issue). 

On the Killing horizon in the background spacetime 
${}^{0}\!g_{\mu \nu}$, the timelike Killing vector 
$\xi_{\scriptscriptstyle (h)}^{\mu}$ becomes null and 
is normal (and so tangent) to the horizon.  
Thus, $\xi_{\scriptscriptstyle (h)}^{\mu}$ satisfies 
\begin{eqnarray} 
{\cal L}_{\xi_{\scriptscriptstyle (h)}} {}^{0}\!g_{\mu \nu} 
\! & \! \! \! = \! \! \! & \! 0 \; , 
\label{eqn:KillingEq} \\ 
{}^{0}\!g_{\mu \nu} \: \xi_{\scriptscriptstyle (h)}^{\mu} \:  
\xi_{\scriptscriptstyle (h)}^{\nu} 
\! & \! \! \! = \! \! \! & \! {\cal O}(v) \; .  
\label{eqn:NullCond} 
\end{eqnarray} 
In addition, it follows that $\xi_{\scriptscriptstyle (h)}^{\mu}$ is 
hypersurface orthogonal\footnote{
In general, the hypersurface orthogonality holds 
only on the horizon. However, in the case of static and 
spherically symmetric spacetimes, it is globally valid.} 
and tangent to the null geodesics 
(parametrized by non-affine parameters) that generate the horizon, 
\begin{eqnarray} 
\xi_{\scriptscriptstyle (h)}^{[ \mu} \, {}^{0} \nabla^{\nu} \, 
\xi_{\scriptscriptstyle (h)}^{\rho ]} 
\! & \! \! \! = \! \! \! & \! 0 \; , 
\label{eqn:HyperOrthog} \\
\xi_{\scriptscriptstyle (h)}^{\nu} {}^{0} \nabla_{\nu} \,  
\xi_{\scriptscriptstyle (h)}^{\mu} 
\! & \! \! \! = \! \! \! & \! \kappa_0 \: 
\xi_{\scriptscriptstyle (h)}^{\mu} + {\cal O}(v) \; ,   
\label{eqn:KappaDef}
\end{eqnarray} 
where ${}^{0} \nabla_{\mu}$ denotes the covariant 
derivative associated with the background metric ${}^{0}\!g_{\mu \nu}$. 
The temperature of the horizon is then given as 
$T = \kappa_0 \, / \, 2 \pi$ by the surface gravity $\kappa_0$. 

It is important for a later discussion 
to recall that the absolute value of the surface 
gravity depends on the normalization of the Killing vector 
$\xi^{\mu}_{\scriptscriptstyle (h)}$, which we can see from 
Eq.(\ref{eqn:KappaDef}). 
In the usual circumstances, the normalization of 
$\xi^{\mu}_{\scriptscriptstyle (h)}$ is 
determined at a far region away from the horizon. 
It is normalized at the center in the case of the de Sitter 
spacetime, and at infinity in the case of the Schwarzschild spacetime. 
Then, in the de Sitter spacetime, $\xi^{\mu}_{\scriptscriptstyle (h)}$ is given by 
\begin{equation} 
\xi_{\scriptscriptstyle (h)}^{\mu} = - \; \kappa_0 \: v \: 
\frac{\partial}{\partial v} 
+ \kappa_0 \: u \: \frac{\partial}{\partial u} \; , 
\label{eqn:XitNorm} 
\end{equation} 
with $\kappa_0 = 1 / \ell$, while $u$ and $v$ are interchanged and 
$\kappa_0 = 1 / 4 M$ in the case of the Schwarzschild spacetime. 

There are another Killing vectors $\xi^{\mu}_{\scriptscriptstyle (i)}$ 
in spherically symmetric spacetimes,  
which generate rigid $O(3)$ rotations of a two-sphere. 
The Lie brackets of those Killing vectors are given by 
\begin{eqnarray} 
\bigl[ \xi_{\scriptscriptstyle (i)} , \xi_{\scriptscriptstyle (j)} 
\bigr]^{\mu} \! & \! \! \! = \! \! \! & \! 
{\cal L}_{\xi_{\scriptscriptstyle (i)}} \xi^{\mu}_{\scriptscriptstyle (j)} 
= C_{\scriptscriptstyle (i) (j)}^{\scriptscriptstyle ~~~~ (k)} \: 
\xi^{\mu}_{\scriptscriptstyle (k)} \; , 
\label{eqn:CommutO3} \\ 
\bigl[ \xi_{\scriptscriptstyle (h)} , \xi_{\scriptscriptstyle (j)} 
\bigr]^{\mu} \! & \! \! \! = \! \! \! & \! 
{\cal L}_{\xi_{\scriptscriptstyle (h)}} \xi^{\mu}_{\scriptscriptstyle (i)} 
= 0 \; , 
\label{eqn:CommutXiHO3} 
\end{eqnarray} 
where 
$C_{\scriptscriptstyle (i) (j)}^{\scriptscriptstyle ~~~~ (k)}$ is 
the structure constant of the $O(3)$ group. 

In what follows, we will present explicit results only in the case 
of the future cosmological horizon in the de Sitter spacetime, 
which is located at $v = 0$. However, it will be clear that 
the analyses in this section, except for the last subsection, 
and those in the next section can be readily applied 
to arbitrary Killing horizons 
with the non-vanishing surface gravity in static and spherically 
symmetric spacetimes.

\subsection{Asymptotic Killing vectors} 
\label{subsec:symmetry} 

Now we analyze asymptotic symmetries on Killing horizons and 
the group associated with those symmetries. 
Since Killing horizons are null hypersurfaces, 
it is expected that the group of asymptotic symmetries on Killing 
horizons will shear some features with the B.M.S. group\cite{BMS}, 
which is the subgroup of diffeomorphisms that preserve the 
asymptotic form of the metric at the null infinity in an 
asymptotically flat spacetime. 
Then, we will refer to the features of the B.M.S. group in order to
define and analyze asymptotic symmetries on Killing horizons.  

In deriving the B.M.S. group, we consider metric 
perturbations on the flat background spacetime, 
which fall off sufficiently at the null infinity. 
Similarly, here we perturb the metric ${}^{0}g_{\mu \nu}$ of 
the background spacetime, 
so that the perturbed metrics $g_{\mu \nu}$ obey the 
boundary condition that they coincide with ${}^{0}g_{\mu \nu}$ on 
the Killing horizon at $v = 0$ in the background spacetime. 
However, in analyses of asymptotic symmetries, we need not 
consider the whole of the horizon. Instead, we are concerned only with 
a local region of the Killing horizon spanned by a horizon two-sphere 
(an intersection of the horizon with a Cauchy surface) during a small 
interval of a null affine parameter on the horizon. 
Then, by assuming suitable differentiability\footnote{
It is sufficient to assume that the perturbed metrics $g_{\mu \nu}$ are 
$C^1$ tensors in this subsection, and $C^2$ tensors 
in the next subsection, 
where we will consider reduced asymptotic symmetries. 
In order to derive asymptotic Killing vectors, however, 
we can assume that $g_{\mu \nu}$ are analytic without loss of 
generality, because higher order terms in $v$ do not 
make any contribution. 
} 
with respect to regular coordinates 
in the local region on the horizon, 
the perturbed metrics $g_{\mu \nu}$ will be written near the horizon as 
\begin{equation}
g_{\mu \nu} = {}^{0}\!g_{\mu \nu} 
+ v \, g'_{\mu \nu}(u, \theta, \phi) 
+ {\cal O}(v^2) 
= {}^{0}\!g_{\mu \nu} + {\cal O}(v) \; ,   
\label{eqn:LimitMetricDS}
\end{equation} 
where the perturbations of the metric 
$g'_{\mu \nu}(u, \theta, \phi)$ 
depend arbitrarily on $u$, $\theta$ and $\phi$. 

If we focus only on the horizon, the metrics with the form of 
Eq.(\ref{eqn:LimitMetricDS}) cannot be distinguished from each other. 
Then, it will be natural to define an equivalence class, 
where all the metrics subject to the boundary condition 
Eq.(\ref{eqn:LimitMetricDS}) are regarded as  
equivalent. We also consider the coordinate transformations 
which transform any metric in the equivalence class 
into another (possibly the same) one in the equivalence class. 
Thus, we obtain the symmetries between the 
metrics in the equivalence class and the subgroup of diffeomorphisms, 
which define the asymptotic symmetries on Killing horizons and 
the asymptotic symmetry group, respectively, 
in a similar manner to the B.M.S. group. 

The infinitesimal transformations of the asymptotic symmetry group 
are then generated by the infinitesimal coordinate transformations 
$x^{\mu} \rightarrow x^{\mu} - \zeta^{\mu}$, where the metric 
$g_{\mu \nu}$ is transformed by the Lie derivatives along 
$\zeta^{\mu}$, as 
\begin{equation} 
g_{\mu \nu} ~~ \longrightarrow ~~ \overline{g}_{\mu \nu} = g_{\mu \nu} 
+ {\cal L}_{\zeta} g_{\mu \nu} \; , 
\end{equation} 
with $g_{\mu \nu}$ and $\overline{g}_{\mu \nu}$ satisfying 
the boundary condition near the horizon, i.e.,  
$\overline{g}_{\mu \nu} = g_{\mu \nu} + {\cal O}(v) 
= {}^{0}\!g_{\mu \nu} + {\cal O}(v)$. 
Therefore, those transformations are described by 
{\it asymptotic Killing vectors} $\zeta^{\mu}$, which 
satisfy the asymptotic Killing equation   
\begin{equation}
{\cal L}_{\zeta} \: g_{\mu \nu} = {\cal O}(v) \; ,  
\label{eqn:AsymKillingAbs} 
\end{equation} 
for arbitrary forms of the perturbed metric $g_{\mu \nu}$ subject to 
the boundary condition Eq.(\ref{eqn:LimitMetricDS}). 

By writing the form of asymptotic Killing vectors $\zeta^{\mu}$ 
near the horizon as 
\begin{equation}
\zeta^{\mu} = 
\zeta^{\mu}_{\scriptscriptstyle (0)}(u,\theta,\phi) + 
v \; \zeta^{\mu}_{\scriptscriptstyle (1)}(u,\theta,\phi) 
+ {\cal O}(v^{2}) \; , 
\label{eqn:ZetaExpand} 
\end{equation} 
and requiring that the asymptotic Killing equation 
Eq.(\ref{eqn:AsymKillingAbs}) is satisfied by arbitrary forms of 
$g'_{\mu \nu}$, we obtain a set of equations for the components of 
$\zeta^{\mu}$. The explicit forms of those equations are shown in 
Appendix \ref{sec:AsymKillEq}. A remarkable feature is that 
the subset of the equations which govern 
$\zeta^{\theta}_{\scriptscriptstyle (0)}$ and 
$\zeta^{\phi}_{\scriptscriptstyle (0)}$ decouples from the rest 
and gives exactly the Killing equation on a two-sphere, while the 
other equations determine the forms of 
$\zeta^{u}_{\scriptscriptstyle (1)}$, 
$\zeta^{v}_{\scriptscriptstyle (1)}$, 
$\zeta^{\theta}_{\scriptscriptstyle (1)}$ and 
$\zeta^{\phi}_{\scriptscriptstyle (1)}$ in terms of the undetermined 
function $\zeta^{u}_{\scriptscriptstyle (0)}$. When we set as 
\begin{equation} 
\omega(u,\theta,\phi) \equiv \kappa_0 \: u \: 
\zeta^{u}_{\scriptscriptstyle (0)}\!(u,\theta,\phi) \; , 
\label{eqn:omegafuncDef} 
\end{equation} 
the general form of asymptotic Killing vectors $\zeta^{\mu}$ is 
given, by using the Killing vectors $\xi^{\mu}_{\scriptscriptstyle (h)}$ 
and $\xi^{\mu}_{\scriptscriptstyle (i)}$, as 
\begin{equation}
\zeta^{\mu} = \omega(u,\theta,\phi) \; \xi^{\mu}_{\scriptscriptstyle (h)} 
+ a^{\scriptscriptstyle (i)} \, \xi^{\mu}_{\scriptscriptstyle (i)} 
+ v \, X^{\mu}(\omega) + v^2 \, Y^{\mu} + {\cal O}(v^3) \; , 
\label{eqn:ZetaCom} 
\end{equation}
where we defined $X^{\mu}(\omega)$ as 
\begin{equation} 
X^{\mu}(\omega) \equiv 2 \: \kappa_0 \, u \, \nabla^{\mu} 
\omega(u,\theta,\phi) \; , 
\label{eqn:XDef}
\end{equation} 
without loss of generality. Whereas other forms of 
$X^{\mu}(\omega)$ different from Eq.(\ref{eqn:XDef}) 
by terms of ${\cal O}(v)$ are possible, such difference can be absorbed into the term of ${\cal O}(v^2)$.     
The coefficients $a^{\scriptscriptstyle (i)}$ are arbitrary constants, 
where summation over the index $(i)$ of the $O(3)$ group is understood, 
and $Y^{\mu}$ is not determined by the asymptotic Killing equation. 

The first term in Eq.(\ref{eqn:ZetaCom}) generates 
position-dependent translations in the null direction on the horizon, 
which are described by the arbitrary function $\omega(u, \theta, \phi)$ 
and referred to as {\it supertranslations}. 
Those are similar to supertranslations in the B.M.S. group, 
but the latter depend only on the angular coordinates. 
It is a distinguishing feature that 
supertranslations of the asymptotic symmetries on Killing horizons 
can depend also on the null coordinate $u$ in arbitrary manners, 
and it results in interesting consequences, as we will see later. 
The second term represents rigid $O(3)$ rotations of 
a horizon two-sphere. 
Thus, we find that the asymptotic symmetry group consists of 
$O(3)$ rotations of a horizon two-sphere and supertranslations along 
the null direction on the horizon. 
Then, in terms of $\omega(u, \theta, \phi)$, 
the asymptotic Killing equation is described as 
\begin{equation} 
\xi^{( \nu}_{\scriptscriptstyle (h)} \nabla^{\mu )} \omega 
+ X^{( \nu}(\omega) \: \nabla^{\mu )} v  
= {\cal O}(v) \; . 
\label{eqn:AsymKillingEqOmega} 
\end{equation} 

We notice that since the hypersurface $v = 0$ is null even in 
the perturbed spacetimes, normal vectors 
to that hypersurface are also null and proportional to 
$\xi^{\mu}_{\scriptscriptstyle (h)}$. 
In fact, $\nabla^{\mu} v$ is written as 
\begin{equation}
\nabla^{\mu} v = - \, \frac{1}{2 \: \kappa_0 \, u} \; 
\xi^{\mu}_{\scriptscriptstyle (h)} + {\cal O}(v) \; . 
\label{eqn:XiDvRelation}
\end{equation} 
The inner products of the Killing vectors with  
$\nabla_{\mu} v$, as well as those with $\nabla_{\mu} u$, 
are then given by 
\begin{eqnarray}
\xi^{\mu}_{\scriptscriptstyle (h)} \nabla_{\mu} v = 
- \: \kappa_0 \: v = {\cal O}(v) \; , ~~~~~~ 
\xi^{\mu}_{\scriptscriptstyle (i)} \nabla_{\mu} v \! 
& \! \! \! = \! \! \! & \! 0 \label{eqn:InProV} \\ 
\xi^{\mu}_{\scriptscriptstyle (h)} \nabla_{\mu} u = 
\: \kappa_0 \: u \; , ~~~~~~ 
\xi^{\mu}_{\scriptscriptstyle (i)} \nabla_{\mu} u \! 
& \! \! \! = \! \! \! & \! 0 
\; ,  \label{eqn:InProU}  
\end{eqnarray} 
and hence, particularly, 
\begin{equation} 
\zeta^{\mu} \: \nabla_{\mu} v = {\cal O}(v) \; . 
\label{eqn:TangentZeta} 
\end{equation} 
Therefore, asymptotic Killing vectors $\zeta^{\mu}$ 
are always tangent to the null hypersurface $v = 0$ 
even in the perturbed spacetimes, and thus they never push or 
pull that hypersurface. 

The Lie bracket between two asymptotic Killing vectors,
\begin{equation}
\zeta^{\mu}_{1} = \omega_{1} \; 
\xi^{\mu}_{\scriptscriptstyle (h)} 
+ a^{\scriptscriptstyle (i)}_{1} \: 
\xi^{\mu}_{\scriptscriptstyle (i)} + v \: X^{\mu}(\omega_{1}) 
+ {\cal O}(v^2) \; , ~~~~ 
\zeta^{\mu}_{2} = \omega_{2} \; 
\xi^{\mu}_{\scriptscriptstyle (h)} 
+ a^{\scriptscriptstyle (i)}_{2} \: 
\xi^{\mu}_{\scriptscriptstyle (i)} + v \: X^{\mu}(\omega_{2}) 
+ {\cal O}(v^2) \; , 
\end{equation} 
is calculated, by using Eqs.(\ref{eqn:XDef})--(\ref{eqn:InProU}) 
as well as Eqs.(\ref{eqn:CommutO3}) and (\ref{eqn:CommutXiHO3}), as 
\begin{equation} 
\bigl[ \zeta_{1} , \zeta_{2} \bigr]^{\mu} = 
{\cal L}_{\zeta_{1}} \zeta^{\mu}_{2} = \omega_{12} \: 
\xi^{\mu}_{\scriptscriptstyle (h)} 
+ a^{\scriptscriptstyle (i)}_{12} \: \xi^{\mu}_{\scriptscriptstyle (i)} 
+ v \: X^{\mu}(\omega_{12}) + {\cal O}(v^2) 
\; , \label{eqn:AsymKillingCommut} 
\end{equation} 
where $\omega_{12} \equiv 
\zeta^{\nu}_{1} \, \nabla_{\nu} \, \omega_{2} - 
\zeta^{\nu}_{2} \, \nabla_{\nu} \, \omega_{1}$ 
and $a^{\scriptscriptstyle (i)}_{12} \equiv 
a^{\scriptscriptstyle (j)}_{1} \, a^{\scriptscriptstyle (k)}_{2} \, 
C_{\scriptscriptstyle (j) (k)}^{\scriptscriptstyle ~~~~ (i)}$. 
We see that 
$\bigl[ \zeta_{1} , \zeta_{2} \bigr]^{\mu}$ 
has the form of an asymptotic 
Killing vector. It explicitly shows 
that the transformations generated by the Lie derivatives along 
asymptotic Killing vectors $\zeta^{\mu}$ close, 
and they indeed form the Lie bracket algebra of the asymptotic symmetry 
group, as they should.  

We also note here that the Lie bracket algebra of 
the asymptotic symmetry group 
contains the $\mathit{Diff}(S^1)$ subalgebra, 
which is necessary, but not sufficient, in order to derive 
the Bekenstein--Hawking formula by 
the same method as that of the B.T.Z. black hole. 
It is easy to see from Eq.(\ref{eqn:AsymKillingCommut}) 
that the supertranslations dependent only on the null coordinate, i.e., 
the asymptotic Killing vectors with $\omega = \omega(u)$ and 
$a^{\scriptscriptstyle (i)} = 0$, form a subalgebra, and commute with 
the isometries of $O(3)$ rotations. 
In order to analyze this subalgebra, we decompose $\omega(u)$ 
into the Fourier modes associated with 
the translational invariance along the null direction on 
the Killing horizon in the background spacetime. 
Then, we write as 
\begin{equation} 
\omega(u) = \int \! dk \: 
\frac{\varpi_{k}}{\kappa_0} \: 
\exp \! \left[ i \: k \: \kappa_0 \: U \right] 
= \int \! dk \: 
\frac{\varpi_{k}}{\kappa_0} \: 
\left( \kappa_0 \: u \right)^{i k}
\; , 
\label{eqn:SuperTransDecomp} 
\end{equation} 
where we introduced the infinitesimal dimensionless coefficients 
$\varpi_{k}$. 
The Fourier modes are defined with respect to the null 
coordinate $U$, not the Kruskal coordinate 
$u$, because the Killing vector $\xi^{\mu}_{\scriptscriptstyle (h)}$ is 
written on the horizon ($v = 0$) as 
\begin{equation} 
\xi^{\mu}_{\scriptscriptstyle (h)} 
=  \kappa_0 \: u \: \frac{\partial}{\partial u} 
= \frac{\partial}{\partial U} 
\; . 
\end{equation} 
If we define $\zeta^{\mu}_k$ as 
\begin{equation} 
\zeta^{\mu}_k \equiv \frac{1}{\kappa_0} \: 
\left( \kappa_0 \: u \right)^{i k} \xi^{\mu}_{\scriptscriptstyle (h)} \; , 
\label{eqn:SuperTransNull} 
\end{equation} 
for each Fourier mode, the Lie bracket algebra of $\zeta^{\mu}_k$ 
is given by  
\begin{equation} 
\bigl[ \zeta_k , \zeta_{k'} \bigr]^{\mu} = i \: ( k' - k ) \; 
\zeta^{\mu}_{k + k'} 
\; . \label{eqn:DiffS1} 
\end{equation} 
Since we are focusing on the local (finite) region on the horizon with 
translational invariance, it will be natural to impose a periodic boundary 
condition on $\omega(u)$ along the null direction. 
In this case, $k$ will take discrete values, and then 
Eq.(\ref{eqn:DiffS1}) is isomorphic to the $\mathit{Diff}(S^1)$ algebra, 
where we appropriately rescale $\zeta^{\mu}_k$, if necessary. 
Therefore, the asymptotic symmetry group on Killing horizons 
is reduced to the $\mathit{Diff}(S^1)$ subgroup in a natural manner, 
by suppressing the dependence of supertranslations on 
the angular coordinates of the spherically symmetric Killing horizon. 

\subsection{Reduction of the asymptotic symmetry group} 
\label{subsec:Reduction} 

As we derived above, 
the asymptotic symmetry group on Killing horizons contains 
supertranslations, similarly to the B.M.S. group. 
It is known that the B.M.S. group reduces to the Poincar\'e group, 
which does not contain supertranslations, when we impose 
the boundary condition stronger 
than that imposed to derive the B.M.S. group. Here, we consider 
reduction of the asymptotic symmetry group on Killing horizons 
in the same manner. 
It will be helpful in understanding the structure of 
the asymptotic symmetry group. 

We thus derive reduced asymptotic symmetries in a similar way 
to the previous subsection, but with a stringent boundary 
condition on the asymptotic form of the metric. 
Perturbations of the metric are now required to be 
${\cal O}(v^2)$, as 
\begin{equation} 
g_{\mu \nu} = {}^{0}\!g_{\mu \nu} 
+ \frac{1}{2} \: v^2 g''_{\mu \nu}(u, \theta, \phi) 
+ {\cal O}(v^3) \; ,    
\label{eqn:LimitMetricDSSecond} 
\end{equation} 
and remain ${\cal O}(v^2)$ when the metric is transformed 
by the Lie derivatives along {\it reduced asymptotic Killing vectors} 
$\eta^{\mu}$, as 
$g_{\mu \nu} \rightarrow g_{\mu \nu} + {\cal L}_{\eta} g_{\mu \nu}$. 
Hence, we demand that the asymptotic Killing equation of second order  
\begin{equation} 
{\cal L}_{\eta} g_{\mu \nu} = {\cal O}(v^2) \; , 
\label{eqn:AsymKillingSecond} 
\end{equation} 
holds for arbitrary perturbations of the metric $g''_{\mu \nu}$. 

However, the term of ${\cal O}(v)$ in $g_{\mu \nu}$ affects 
the forms of reduced asymptotic Killing vectors $\eta^{\mu}$, 
and then $\eta^{\mu}$ are not determined unless the background metric 
${}^{0}\!g_{\mu \nu}$ is specified. 
Here, we specifically consider the de Sitter spacetime, 
which is maximally symmetric and possesses a Killing horizon. 
This is because here we attempt to compare 
the asymptotic symmetry group on the Killing horizon 
with the isometry group in the maximally symmetric spacetime 
in order to understand the structure of the asymptotic symmetry group. 
Thus, in this subsection, 
we focus on the asymptotic symmetry group and its reduction on 
the future cosmological horizon 
in the de Sitter spacetime, and compare the reduced asymptotic 
symmetry group with the de Sitter group. 

We solve Eq.(\ref{eqn:AsymKillingSecond}), by writing the 
form of reduced asymptotic Killing vectors as 
\begin{equation}
\eta^{\mu} = 
\eta^{\mu}_{\scriptscriptstyle (0)}(u,\theta,\phi) + 
v \; \eta^{\mu}_{\scriptscriptstyle (1)}(u,\theta,\phi) 
+ v^2 \; \eta^{\mu}_{\scriptscriptstyle (2)}(u,\theta,\phi) 
+ {\cal O}(v^{3}) \; , 
\label{eqn:EtaExpand} 
\end{equation} 
and substituting Eqs.(\ref{eqn:DSKrsMetric}) and (\ref{eqn:EtaExpand}) 
into Eq.(\ref{eqn:AsymKillingSecond}) 
(see Appendix \ref{sec:AsymKillEq}). 
Then, we find that the solutions of $\eta^{\mu}$ are given by 
linear combinations of the five vectors, which are written as 
\begin{eqnarray} 
\eta^{\mu}_{\scriptscriptstyle (t)} \! & \! \! \! = \! \! \! & \! 
- \; v \; \frac{\partial}{\partial v} 
+ u \; \frac{\partial}{\partial u} + {\cal O}(v^3) \; , 
\label{eqn:AsymKillVecSec01} \\ 
\eta^{\mu}_{\scriptscriptstyle (\parallel)} \! & \! \! \! = \! \! \! & \!  
\frac{v^2}{\ell} \cos \theta \:  
\frac{\partial}{\partial v} 
- \ell \, \cos \theta \: \frac{\partial}{\partial u} 
+ \left( \frac{2}{\ell} \: v - \frac{2 u}{\ell^3} \: v^2 \right) 
\frac{\partial}{\partial \theta}  + {\cal O}(v^3) \; , 
\label{eqn:AsymKillVecSecm2} \\ 
\eta^{\mu}_{\scriptscriptstyle (\parallel\pm)} \! & \! \! \! = \! \! \! & \!  
\frac{v^2}{\ell} 
\sin \theta \; e^{\pm i \phi} \frac{\partial}{\partial v} 
- \ell \: \sin \theta \; e^{\pm i \phi} \frac{\partial}{\partial u} 
- \left( \frac{2}{\ell} \: v - \frac{2 u}{\ell^3} \: v^2 \right) 
\cos \theta \; e^{\pm i \phi} \frac{\partial}{\partial \theta} 
\nonumber \\ 
& & \mp \; i \left( \frac{2}{\ell} \: v - \frac{2 u}{\ell^3} \: v^2 \right) 
\frac{e^{\pm i \phi}}{\sin \theta} \; \frac{\partial}{\partial \phi} 
+ {\cal O}(v^3) \; , 
\label{eqn:AsymKillVecSecmpm} \\ 
\eta^{\mu}_{\scriptscriptstyle (\pm)} \! & \! \! \! = \! \! \! & \!  
e^{\pm i\phi} 
\frac{\partial}{\partial \theta} 
\pm i \; \cot \theta \; e^{\pm i \phi} \frac{\partial}{\partial \phi} 
+ {\cal O}(v^3) \; , 
\label{eqn:AsymKillVecSec2pm} \\ 
\eta^{\mu}_{\scriptscriptstyle (\phi)} \! & \! \! \! = \! \! \! & \!  
\frac{\partial}{\partial \phi} 
+ {\cal O}(v^3) \; . 
\label{eqn:AsymKillVecSec34} 
\end{eqnarray}  
We see that supertranslations are not allowed as reduced asymptotic 
symmetries, as expected. In addition, we also notice, 
by comparing with the exact Killing vectors in the de Sitter spacetime 
(those are listed in Appendix \ref{sec:KillingList}), 
that reduced asymptotic Killing vectors are 
given only by the asymptotic forms of the Killing vectors 
that are tangent to the future cosmological horizon. 

It will be worth while mentioning here the remarkable difference 
between the structure of the B.M.S. group and that of 
the asymptotic symmetry group on the 
cosmological horizon in the de Sitter spacetime. 
It is known that all the exact Killing vectors in the Minkowski spacetime 
have the form of asymptotic Killing vectors of 
the B.M.S. group\cite{BMS}. 
In other words, the B.M.S. group contains the Poincar\'e group. 
However, in the case of the 
asymptotic symmetry group on the cosmological horizon, 
some Killing vectors in the de Sitter spacetime 
do not belong to asymptotic Killing vectors. 
In fact, we see that $\xi^{\mu}_{\scriptscriptstyle (\bot)}$ and 
$\xi^{\mu}_{\scriptscriptstyle (\bot\pm)}$, given by 
Eqs.(\ref{eqn:KillingDSp2}) 
and (\ref{eqn:KillingDSppm}) in Appendix \ref{sec:KillingList}, 
are not tangent to the horizon and 
do not have the form of asymptotic 
Killing vectors. 
Correspondingly, those vectors do not belong to reduced asymptotic 
Killing vectors either. 
Thus, we find that the asymptotic symmetry group 
does not contain the de Sitter group. 
The full isometries of the de Sitter group will not be recovered 
unless we specify the complete form of the metric, 
not only its asymptotic form on the horizon.
This difference between the asymptotic symmetry 
group on the cosmological horizon and the B.M.S. group will be  
a consequence of the fact that 
the cosmological horizon is located at a finite affine distance from 
the center of the de Sitter spacetime, 
whereas the null infinity is infinitely far 
from any point in the spacetime. 
On the other hand, we also note that supertranslations belong to 
the asymptotic symmetry group, but do not 
to the de Sitter group or the reduced asymptotic symmetry group, 
similarly to those in the B.M.S. group.

\section{Asymptotic Killing horizons} 
\label{sec:AsymptoticKillingHorizon} 
\eqnum{0} 

The asymptotic symmetry group we have derived in the previous section 
leaves invariant the form of the metric on the Killing horizon in the 
background spacetime. 
However, it may not necessarily indicate that 
local structures of the horizon are also unaffected. 
In this section, we analyze how transformations generated by 
the asymptotic symmetry group affect local properties 
that characterize the structures of Killing horizons. 

\subsection{Local properties of Killing horizons} 

We first specify local properties of Killing horizons. 
On the Killing horizon in the background spacetime 
${}^{0}\!g_{\mu \nu}$, the timelike 
Killing vector $\xi^{\mu}_{\scriptscriptstyle (h)}$ satisfies 
the properties Eq.(\ref{eqn:KillingEq})--(\ref{eqn:KappaDef}). 
Among them, the Killing equation Eq.(\ref{eqn:KillingEq}) and 
the hypersurface orthogonality Eq.(\ref{eqn:HyperOrthog}) hold 
globally in the spacetime. 
However, if we focus only on the horizon, 
we need not take care that those properties are satisfied 
in far regions away from the horizon. 
We only have to ensure that the asymptotic forms of 
those properties are satisfied on the horizon. 
In addition, from the standpoint in the previous section, 
where all the metrics subject to the boundary condition 
Eq.(\ref{eqn:LimitMetricDS}) are regarded as equivalent, 
it is natural to treat the perturbed metrics $g_{\mu \nu}$ 
on an equal footing with the background metric ${}^{0}\!g_{\mu \nu}$. 
We are thus led to an extended notion as follows. 

Under the presence of the exact Killing horizon 
($v = 0$) generated by the Killing vector 
$\xi^{\mu}_{\scriptscriptstyle (h)}$ in the background spacetime 
${}^{0}\!g_{\mu \nu}$, we define an {\it asymptotic Killing horizon} 
as the region of the hypersurface $v = 0$ in a perturbed spacetime 
$g_{\mu \nu}$, over which the same vector 
$\xi^{\mu}_{\scriptscriptstyle (h)}$ as the Killing vector 
in the background spacetime satisfies 
\begin{eqnarray} 
{\cal L}_{\xi_{\scriptscriptstyle (h)}} g_{\mu \nu} 
\! & \! \! \! = \! \! \! & \! {\cal O}(v) \; , 
\label{eqn:AsyKillingEq} \\ 
g_{\mu \nu} \: \xi_{\scriptscriptstyle (h)}^{\mu} \:  
\xi_{\scriptscriptstyle (h)}^{\nu} 
\! & \! \! \! = \! \! \! & \! {\cal O}(v) \; ,  
\label{eqn:AsyNullCond} \\
\xi_{\scriptscriptstyle (h)}^{[ \mu} \nabla^{\nu} 
\xi_{\scriptscriptstyle (h)}^{\rho ]} 
\! & \! \! \! = \! \! \! & \! {\cal O}(v) \; , 
\label{eqn:AsyHyperOrthog} \\
\xi_{\scriptscriptstyle (h)}^{\nu} \nabla_{\nu} \,  
\xi_{\scriptscriptstyle (h)}^{\mu} 
\! & \! \! \! = \! \! \! & \! \kappa \; 
\xi_{\scriptscriptstyle (h)}^{\mu} + {\cal O}(v) \; .   
\label{eqn:AsyKappaDef}
\end{eqnarray} 
Then, an asymptotic Killing horizon has the same local 
properties as those of the exact Killing horizon, and hence 
it locally looks like a part of a Killing horizon 
when we focus only on the horizon. 
We see from the boundary condition Eq.(\ref{eqn:LimitMetricDS}) that 
an asymptotic Killing horizon is indeed a null hypersurface. 
In addition, Eqs.(\ref{eqn:AsyKillingEq}) and (\ref{eqn:AsyNullCond}) 
imply that $\xi^{\mu}_{\scriptscriptstyle (h)}$ is null and 
plays the role of a Killing vector only on an asymptotic Killing horizon, 
which we will call the {\it horizon generating vector} in what follows. 
Hence, a photon traveling on an asymptotic Killing horizon will feel 
like trapped at ``places of the same geometry'', in a similar sense to 
that a photon moving along the exact Killing horizon 
stays at $r = r_{\scriptscriptstyle H}$. 
We define the surface gravity $\kappa$ of an asymptotic 
Killing horizon by Eq.(\ref{eqn:AsyKappaDef}), 
similarly to the case of the exact Killing horizon, 
while Eq.(\ref{eqn:AsyHyperOrthog}), 
together with Eq.(\ref{eqn:AsyKillingEq}), serves to ensure 
compatibility of Eq.(\ref{eqn:AsyKappaDef}) with other possible 
definitions of the surface gravity\footnote{In the case of the exact 
Killing horizon, 
the definitions $\nabla_{\mu} ( \xi^{\alpha}_{\scriptscriptstyle (h)} 
\xi_{{\scriptscriptstyle (h)} \alpha} ) = 
- \, \kappa \: \xi_{{\scriptscriptstyle (h)} \mu}$ and 
$\kappa^2 = - \, ( \nabla^{\alpha} \xi^{\beta}_{\scriptscriptstyle (h)} ) \, 
( \nabla_{\alpha} \xi_{{\scriptscriptstyle (h)} \beta} ) / 2$ give 
the same value of the surface gravity as that defined by 
Eq.(\ref{eqn:KappaDef}). 
Those definitions of the surface gravity result  
from Eq.(\ref{eqn:KappaDef}), by using Eqs.(\ref{eqn:KillingEq}) 
and (\ref{eqn:HyperOrthog}).}. 

We emphasize here that the surface gravity of an asymptotic Killing 
horizon does not necessarily have the physical meaning 
as the temperature of the horizon. 
This is because we have specified only the local geometry near 
an asymptotic Killing horizon 
in a perturbed spacetime, and hence it is not clear how the surface 
gravity of the asymptotic Killing horizon is related to thermal radiation 
outside the horizon. 
On the contrary, in the original derivation of 
Hawking radiation\cite{HawkingRad}, it is shown that an observer 
{\it at infinity} receives thermal radiation with the 
temperature proportional to the surface gravity. 
This fact can be interpreted as indicating that we need to know 
the global behavior of the spacetime in order to 
define the temperature of a black hole\footnote{
In the Euclidean approach to 
black hole thermodynamics\cite{EuclideanBH}, the temperature of 
a black hole is derived by imposing regular periodicity near 
the horizon in the Euclidean section. 
However, also in this case, we will need to specify 
the global behavior of the Euclidean section, so that the system is 
in thermal equilibrium, i.e., the Hartle--Hawking state.  
In fact, the energy-momentum tensor is singular 
on the horizon, when we consider a quantum field out of equilibrium 
in a static spacetime\cite{BirrellDavies}.}. 
Then, in this paper, we take the point of view that 
the surface gravity of an asymptotic Killing horizon is not 
directly related to the temperature. 
Accordingly, we do not require that the surface gravity of 
an asymptotic Killing horizon 
is constant over the horizon. It does not necessarily violate 
the zero-th law of black hole thermodynamics, because it states 
that the surface gravity of the exact Killing horizon, 
not that of an asymptotic Killing horizon, 
is constant. 

\subsection{Transformations} 

Here, we analyze how the four properties of asymptotic 
Killing horizons Eqs.(\ref{eqn:AsyKillingEq})--(\ref{eqn:AsyKappaDef}) 
are affected under transformations generated by 
the asymptotic symmetry group. 

In the previous section, the asymptotic symmetry group has been derived 
as a subgroup of diffeomorphisms. When we consider infinitesimal 
diffeomorphisms, the metric 
$g_{\mu \nu}$ and the horizon generating vector 
$\xi_{\scriptscriptstyle (h)}^{\mu}$ undergo 
the simultaneous transformations generated by the Lie derivatives 
along asymptotic Killing vectors $\zeta^{\mu}$, as 
\begin{eqnarray} 
g_{\mu \nu} & ~~ \longrightarrow ~~ & \widehat{g}_{\mu \nu} = 
g_{\mu \nu} + {\cal L}_{\zeta} g_{\mu \nu} \; , \\ 
\xi_{\scriptscriptstyle (h)}^{\mu} & ~~ \longrightarrow ~~ & 
\widehat{\xi}_{\scriptscriptstyle (h)}^{\mu} = 
\xi_{\scriptscriptstyle (h)}^{\mu} + {\cal L}_{\zeta} 
\xi_{\scriptscriptstyle (h)}^{\mu} \; . 
\end{eqnarray} 
Since any covariant expressions are transformed by 
the Lie derivatives under infinitesimal diffeomorphisms 
and we have Eq.(\ref{eqn:TangentZeta}), 
we see, by taking the Lie derivatives, that the four properties of asymptotic Killing horizons are transformed as 
\begin{eqnarray} 
{\cal L}_{\xi_{\scriptscriptstyle (h)}} g_{\mu \nu} = {\cal O}(v) 
& ~~ \longrightarrow ~~ & 
{\cal L}_{\hat{\xi}_{\scriptscriptstyle (h)}} \widehat{g}_{\mu \nu} 
= {\cal O}(v) \; , 
\label{eqn:KillingEqDif} \\
g_{\mu \nu} \: \xi_{\scriptscriptstyle (h)}^{\mu} \:  
\xi_{\scriptscriptstyle (h)}^{\nu} = {\cal O}(v) 
& ~~ \longrightarrow ~~ &  
\widehat{g}_{\mu \nu} \: \widehat{\xi}_{\scriptscriptstyle (h)}^{\mu} \: 
\widehat{\xi}_{\scriptscriptstyle (h)}^{\nu} = {\cal O}(v) \; , 
\label{eqn:NullCondDif} \\
\xi_{\scriptscriptstyle (h)}^{[ \mu} \nabla^{\nu}  
\xi_{\scriptscriptstyle (h)}^{\rho ]} = {\cal O}(v) 
& ~~ \longrightarrow ~~ & 
\widehat{\xi}_{\scriptscriptstyle (h)}^{[ \mu} 
\widehat{\nabla}^{\nu} 
\widehat{\xi}_{\scriptscriptstyle (h)}^{\rho ]} = {\cal O}(v) \; , 
\label{eqn:OrthogDif} \\ 
\xi^{\nu}_{\scriptscriptstyle (h)} \nabla_{\nu} \, 
\xi^{\mu}_{\scriptscriptstyle (h)} = \kappa \; 
\xi^{\mu}_{\scriptscriptstyle (h)} + {\cal O}(v) 
& ~~ \longrightarrow ~~ & 
\widehat{\xi}^{\nu}_{\scriptscriptstyle (h)} 
\widehat{\nabla}_{\nu} \, \widehat{\xi}^{\mu}_{\scriptscriptstyle (h)} 
= \widehat{\kappa} \; \widehat{\xi}^{\mu}_{\scriptscriptstyle (h)} 
+ {\cal O}(v) \; , 
\label{eqn:SurGraDif} 
\end{eqnarray} 
where $\widehat{\nabla}_{\mu}$ is the covariant derivative 
associated with $\widehat{g}_{\mu \nu}$ 
and the transformed surface gravity $\hat{\kappa}$ is given by 
\begin{equation} 
\widehat{\kappa} = \kappa + {\cal L}_{\zeta} \kappa \; . 
\end{equation} 
We note here that the surface gravity $\kappa$ changes under 
diffeomorphisms, unless $\kappa$ is constant, 
as in the background spacetime. 
Although changes of the surface gravity result in reparametrizations of null geodesics, 
those do not alter the local structures of asymptotic Killing horizons 
defined by Eqs.(\ref{eqn:AsyKillingEq})--(\ref{eqn:AsyKappaDef}), 
except for the proportional factor $\kappa$ 
in Eq.(\ref{eqn:AsyKappaDef}). 
Thus, here we allow for changes of the surface gravity, 
and Eq.(\ref{eqn:AsyKappaDef}) is interpreted as 
stating that $\xi^{\nu}_{\scriptscriptstyle (h)} \nabla_{\nu} \, 
\xi^{\mu}_{\scriptscriptstyle (h)}$ is proportional to 
$\xi^{\mu}_{\scriptscriptstyle (h)}$. Then, 
the four properties of asymptotic Killing horizons are invariant 
under diffeomorphisms, as they should. 
However, diffeomorphisms are trivial transformations. 
We look at the same horizon in the same spacetime. 
Now we explore the possibility that the local properties of 
asymptotic Killing horizons are preserved 
under non-trivial transformations. 

Even if we transform the metric 
by the Lie derivative, it does not necessarily mean that the 
transformation under consideration is a diffeomorphism. 
The metric $g_{\mu \nu}$ (a dynamical field variable) and 
the horizon generating vector $\xi_{\scriptscriptstyle (h)}^{\mu}$ 
(a non-dynamical variable independent of the metric) can be 
separately transformed, in general. This will be illustrated clearly, 
when we  consider the transformation described by 
\begin{eqnarray}
g_{\mu \nu} & ~~ \longrightarrow ~~ & \overline{g}_{\mu \nu} 
= g_{\mu \nu} + {\cal L}_{\zeta} \, g_{\mu \nu} 
\equiv g_{\mu \nu} + v \, q_{\mu \nu} 
+ {\cal O}(v^2) \; , 
\label{eqn:Trans1G}
\\ 
\xi^{\mu}_{\scriptscriptstyle (h)} & ~~ \longrightarrow ~~ & 
\overline{\xi}^{\mu}_{\scriptscriptstyle (h)} 
= \xi^{\mu}_{\scriptscriptstyle (h)} \; .  
\label{eqn:Trans1Xi}
\end{eqnarray} 
Here, the metric is transformed by the Lie derivative, but the horizon 
generating vector $\xi^{\mu}_{\scriptscriptstyle (h)}$ is held fixed. 
We call this transformation a {\it dynamical field transformation}. 

When we notice that $\xi^{\mu}_{\scriptscriptstyle (h)}$ belongs to 
asymptotic Killing vectors, 
we immediately see that Eq.(\ref{eqn:AsyKillingEq}) is 
invariant under dynamical field transformations. 
Since  $\overline{g}_{\mu \nu}$ has the asymptotic form of 
Eq.(\ref{eqn:LimitMetricDS}), and hence 
$\xi^{\mu}_{\scriptscriptstyle (h)}$ is null at $v = 0$, 
it is also clear that 
Eq.(\ref{eqn:AsyNullCond}) remains satisfied. In addition, by writing as 
$\overline{\nabla}_{\mu} \xi^{\nu}_{\scriptscriptstyle (h)} 
= \nabla_{\mu} \xi^{\nu}_{\scriptscriptstyle (h)} + 
\xi^{\rho}_{\scriptscriptstyle (h)} \delta \Gamma^{\nu}_{\rho \mu}$, 
where $\overline{\nabla}_{\mu}$ is the covariant derivative associated 
with $\overline{g}_{\mu \nu}$ and 
\begin{equation} 
\delta \Gamma^{\nu}_{\rho \mu} \equiv \frac{1}{2} \, g^{\nu \lambda} 
\left( \nabla_{\mu} {\cal L}_{\zeta} g_{\lambda \rho} 
+ \nabla_{\rho} {\cal L}_{\zeta} g_{\lambda \mu} 
- \nabla_{\lambda} {\cal L}_{\zeta} g_{\mu \rho} \right) \; ,  
\label{eqn:deltaGamma} 
\end{equation}
we can prove that Eqs.(\ref{eqn:AsyHyperOrthog}) and 
(\ref{eqn:AsyKappaDef}) are preserved, while the surface gravity 
changes as $\kappa \rightarrow \overline{\kappa} \equiv \kappa + 
\overline{\delta} \kappa$, with $\overline{\delta} \kappa$ given as 
\begin{equation} 
\overline{\delta} \kappa 
= \frac{1}{4 \: \kappa_0 \, u} \; 
\xi^{\lambda}_{\scriptscriptstyle (h)} \:  
\xi^{\rho}_{\scriptscriptstyle (h)} \: q_{\lambda \rho} \; .  
\label{eqn:DefBarKappa} 
\end{equation} 
Then, we have 
\begin{eqnarray} 
{\cal L}_{\xi_{\scriptscriptstyle (h)}} g_{\mu \nu} = {\cal O}(v) 
& ~~ \longrightarrow ~~ & 
{\cal L}_{\xi_{\scriptscriptstyle (h)}} \overline{g}_{\mu \nu} 
= {\cal O}(v) \; , 
\label{eqn:KillingEqTr} \\
g_{\mu \nu} \: \xi_{\scriptscriptstyle (h)}^{\mu} \:  
\xi_{\scriptscriptstyle (h)}^{\nu} = {\cal O}(v) 
& ~~ \longrightarrow ~~ &  
\overline{g}_{\mu \nu} \: \xi_{\scriptscriptstyle (h)}^{\mu} \:  
\xi_{\scriptscriptstyle (h)}^{\nu} = {\cal O}(v) \; , 
\label{eqn:NullCondTr} \\
\xi_{\scriptscriptstyle (h)}^{[ \mu} \nabla^{\nu}  
\xi_{\scriptscriptstyle (h)}^{\rho ]} = {\cal O}(v) 
& ~~ \longrightarrow ~~ & \xi_{\scriptscriptstyle (h)}^{[ \mu} 
\overline{\nabla}^{\nu} 
\xi_{\scriptscriptstyle (h)}^{\rho ]} = {\cal O}(v) \; , 
\label{eqn:OrthogTrfin} \\ 
\xi^{\nu}_{\scriptscriptstyle (h)} \nabla_{\nu} \, 
\xi^{\mu}_{\scriptscriptstyle (h)} = \kappa \; 
\xi^{\mu}_{\scriptscriptstyle (h)} + {\cal O}(v) 
& ~~ \longrightarrow ~~ & \xi^{\nu}_{\scriptscriptstyle (h)} 
\overline{\nabla}_{\nu} \, \xi^{\mu}_{\scriptscriptstyle (h)} 
= \overline{\kappa} \; \xi^{\mu}_{\scriptscriptstyle (h)} + {\cal O}(v) \; , 
\label{eqn:SurGraTr} 
\end{eqnarray}
and we see that the four properties of asymptotic Killing horizons  
are preserved under dynamical field transformations. 
In particular, the Killing horizon in the background spacetime is 
always an asymptotic Killing horizon in the perturbed spacetime 
$g_{\mu \nu} = {}^{0}\!g_{\mu \nu} + {\cal L}_{\zeta} g_{\mu \nu}$, 
over which the surface gravity is not constant in general, as we see 
from Eq.(\ref{eqn:DefBarKappa}). 

Although a dynamical field transformation changes the metric 
into another one, we can pull it back by a diffeomorphism to 
the original one, since the metric is transformed by the Lie derivative under a dynamical field transformation. 
When we apply a dynamical field transformation followed by 
such a diffeomorphism, 
we obtain the same metric as 
the original one, but the horizon generating vector 
$\xi^{\mu}_{\scriptscriptstyle (h)}$ is now transformed. 
Thus, it will be appropriate to call this transformation 
a {\it horizon deformation}. 
Under a horizon deformation, the metric $g_{\mu \nu}$ is fixed  
and the horizon generating vector 
$\xi^{\mu}_{\scriptscriptstyle (h)}$ is transformed by 
the Lie derivative along $\zeta^{\mu}$, 
\begin{eqnarray}
g_{\mu \nu} & ~~ \longrightarrow ~~ & 
\widetilde{g}_{\mu \nu} = g_{\mu \nu} \; , 
\label{eqn:Trans2G}
\\ 
\xi^{\mu}_{\scriptscriptstyle (h)} & ~~ \longrightarrow ~~ & 
\widetilde{\xi}^{\mu}_{\scriptscriptstyle (h)} 
= \xi^{\mu}_{\scriptscriptstyle (h)} + {\cal L}_{\zeta} 
\xi^{\mu}_{\scriptscriptstyle (h)} \; , 
\label{eqn:Trans2Xi}
\end{eqnarray} 
where ${\cal L}_{\zeta} \xi^{\mu}_{\scriptscriptstyle (h)}$ 
is calculated, by using Eq.(\ref{eqn:AsymKillingCommut}), as  
\begin{equation} 
{\cal L}_{\zeta} \xi^{\mu}_{\scriptscriptstyle (h)} 
\equiv \widetilde{\delta} \xi^{\mu}_{\scriptscriptstyle (h)} 
= - \: \Omega(u, \theta, \phi) \: \xi^{\mu}_{\scriptscriptstyle (h)} 
+ v \, X^{\mu}(- \Omega) + {\cal O}(v^2) \; ,  
\label{eqn:LieBraZetaXit} 
\end{equation} 
and $\Omega(u, \theta, \phi)$ is defined by 
\begin{equation} 
\Omega(u, \theta, \phi) \equiv 
\xi^{\nu}_{\scriptscriptstyle (h)} \nabla_{\nu} \, \omega(u, \theta, \phi) 
\; .  
\label{eqn:CapOmegaDef} 
\end{equation} 
We immediately notice that $\Omega(u, \theta, \phi)$ vanishes, when 
$\omega(u, \theta, \phi)$ does not depend on the null coordinate $u$. 
Hence, the supertranslations that depend on the null coordinate 
play the essential role in horizon deformations. 

Invariance of the properties of asymptotic Killing horizons 
under dynamical field transformations indicates that they are also 
preserved under horizon deformations, because those two types of 
transformations are equivalent to each other up to diffeomorphisms. 
Indeed, as we see from Eq.(\ref{eqn:LieBraZetaXit}), 
$\tilde{\xi}^{\mu}_{\scriptscriptstyle (h)}$ belongs 
to asymptotic Killing vectors and is null at $v = 0$. 
Hence, it is easy to show that 
Eqs.(\ref{eqn:AsyKillingEq}) and (\ref{eqn:AsyNullCond}) are preserved. 
In addition, by using Eqs.(\ref{eqn:XiDvRelation}), 
(\ref{eqn:AsyHyperOrthog}) and (\ref{eqn:LieBraZetaXit}), 
we also obtain $\tilde{\xi}_{\scriptscriptstyle (h)}^{[ \mu} \nabla^{\nu}  
\tilde{\xi}_{\scriptscriptstyle (h)}^{\rho ]} = {\cal O}(v)$. 
Similarly, we find that $\tilde{\xi}_{\scriptscriptstyle (h)}^{\nu} 
\nabla_{\nu} \,  \tilde{\xi}_{\scriptscriptstyle (h)}^{\mu}$ is 
proportional to $\tilde{\xi}_{\scriptscriptstyle (h)}^{\mu}$ at $v = 0$, 
by calculating to first order of infinitesimal transformations as  
\begin{equation} 
\widetilde{\xi}_{\scriptscriptstyle (h)}^{\nu} \nabla_{\nu} \,  
\widetilde{\xi}_{\scriptscriptstyle (h)}^{\mu} = 
\biggl[ \kappa - 2 \: \kappa \, \Omega 
- \xi_{\scriptscriptstyle (h)}^{\nu} \nabla_{\nu} \, \Omega 
\biggr] \, \xi_{\scriptscriptstyle (h)}^{\mu} + {\cal O}(v) \; . 
\end{equation} 
When we write the proportional factor 
$\tilde{\kappa}$ as 
$\tilde{\kappa} \equiv \kappa + \tilde{\delta} \kappa$, 
we have 
\begin{equation} 
\widetilde{\xi}_{\scriptscriptstyle (h)}^{\nu} \nabla_{\nu} \, 
\widetilde{\xi}_{\scriptscriptstyle (h)}^{\mu} 
= 
\widetilde{\kappa} \; 
\widetilde{\xi}_{\scriptscriptstyle (h)}^{\mu} 
= \biggl[ \kappa - \kappa \, \Omega 
+ \widetilde{\delta} \kappa \biggr] \, \xi_{\scriptscriptstyle (h)}^{\mu} 
+ {\cal O}(v) \; , 
\end{equation} 
and thus $\tilde{\delta} \kappa$ is given by 
\begin{equation} 
\widetilde{\delta} \kappa = - \kappa \, \Omega 
- \xi^{\nu}_{\scriptscriptstyle (h)} \nabla_{\nu} \, 
\Omega \; . 
\label{eqn:DelGKappaDef} 
\end{equation} 
The four properties of asymptotic Killing horizons are then 
transformed as 
\begin{eqnarray} 
{\cal L}_{\xi_{\scriptscriptstyle (h)}} g_{\mu \nu} = {\cal O}(v) 
& ~~ \longrightarrow ~~ & 
{\cal L}_{\tilde{\xi}_{\scriptscriptstyle (h)}} g_{\mu \nu} 
= {\cal O}(v) \; , 
\label{eqn:KillingEqTr2} \\ 
g_{\mu \nu} \: \xi_{\scriptscriptstyle (h)}^{\mu} \:  
\xi_{\scriptscriptstyle (h)}^{\nu} = {\cal O}(v) 
& ~~ \longrightarrow ~~ &  
g_{\mu \nu} \: \widetilde{\xi}_{\scriptscriptstyle (h)}^{\mu} \:  
\widetilde{\xi}_{\scriptscriptstyle (h)}^{\nu} = 
{\cal O}(v) \; , 
\label{eqn:NullCondTr2} \\ 
\xi_{\scriptscriptstyle (h)}^{[ \mu} \nabla^{\nu}  
\xi_{\scriptscriptstyle (h)}^{\rho ]} = {\cal O}(v) 
& ~~ \longrightarrow ~~ &  
\widetilde{\xi}_{\scriptscriptstyle (h)}^{[ \mu} \nabla^{\nu} 
\widetilde{\xi}_{\scriptscriptstyle (h)}^{\rho ]} =  {\cal O}(v) \; , 
\label{eqn:HyperOrthogTr2} \\ 
\xi_{\scriptscriptstyle (h)}^{\nu} \nabla_{\nu} \,  
\xi_{\scriptscriptstyle (h)}^{\mu} = \kappa \; 
\xi_{\scriptscriptstyle (h)}^{\mu} 
+ {\cal O}(v) \; 
& ~~ \longrightarrow ~~ &  
\widetilde{\xi}_{\scriptscriptstyle (h)}^{\nu} \nabla_{\nu} \,  
\widetilde{\xi}_{\scriptscriptstyle (h)}^{\mu} = 
\widetilde{\kappa} \; \widetilde{\xi}_{\scriptscriptstyle (h)}^{\mu} 
+ {\cal O}(v) \; ,  
\label{eqn:KappaDefTr2} 
\end{eqnarray} 
and hence those properties are preserved also 
under horizon deformations. 
It implies that the asymptotic symmetry group generates 
deformations of an asymptotic Killing horizon, 
under which the horizon keeps its local structures. 

We point out here 
that arbitrary covariant equations 
which depend only on dynamical field variables 
(the metric in the present case), but not 
on non-dynamical variables (the horizon generating vector 
$\xi^{\mu}_{\scriptscriptstyle (h)}$, for example), are left invariant 
under both dynamical field transformations 
and horizon deformations. 
In particular, the Einstein equation holds in the spacetimes after transformations, if it is satisfied before the transformations.

\subsection{Surface gravity} 
\label{subsec:SurfaceGravity} 

In order to clarify another aspects of asymptotic Killing horizons, 
here we analyze the behavior of the surface gravity under the above 
transformations. 

We first mention the relation between the changes of the surface gravity 
$\overline{\delta} \kappa$ and $\tilde{\delta} \kappa$, given by 
Eq.(\ref{eqn:DefBarKappa}) and Eq.(\ref{eqn:DelGKappaDef}), respectively. 
Since the dynamical field transformation 
and the horizon deformation generated by the same asymptotic Killing 
vector $\zeta^{\mu}$ compose the corresponding diffeomorphism, 
we find that $\overline{\delta} \kappa$ and $\tilde{\delta} \kappa$ 
are related as $\overline{\delta} \kappa + \tilde{\delta} \kappa 
= {\cal L}_{\zeta} \kappa$. 
Thus, when the surface gravity is constant, 
as it is in the background spacetime, 
we have $\overline{\delta} \kappa + \tilde{\delta} \kappa = 0$. 
The dynamical field transformation and the horizon deformation 
give the same magnitude of the change of the surface gravity 
with the opposite sign. 
Then, in order to analyze the behavior of the surface gravity, 
it is convenient to take advantage of horizon deformations, 
where we can work with the fixed metric 
in the background spacetime, 
while the horizon generating vector and the surface gravity are 
transformed by Eq.(\ref{eqn:LieBraZetaXit}) 
and Eq.(\ref{eqn:DelGKappaDef}), respectively. 

If we consider the case where 
$\Omega$ defined by Eq.(\ref{eqn:CapOmegaDef}) 
is constant, both the horizon generating 
vector and the surface gravity are multiplied by $1 - \Omega$ under 
the horizon deformation. We recall here that, in the case of 
the exact Killing horizon, the normalization of the Killing vector 
$\xi^{\mu}_{\scriptscriptstyle (h)}$ is usually fixed by 
the condition imposed far from the horizon.  
However, since we are focusing only on the horizon, there will be 
no reason to persist in such normalization. When we change 
the normalization of $\xi^{\mu}_{\scriptscriptstyle (h)}$, the value 
of the surface gravity also changes by the same factors. 
We see that the horizon deformations with $\Omega = const.$ 
correspond to those trivial transformations, 
which are possible even in the case of the exact Killing horizon. 
In other words, by the re-normalization of the horizon generating vector 
 $\tilde{\xi}^{\mu}_{\scriptscriptstyle (h)} \rightarrow 
( 1 - \Omega )^{-1} \, \tilde{\xi}^{\mu}_{\scriptscriptstyle (h)}$, 
the surface gravity is brought back to the original value $\kappa$. 
In fact, the horizon deformations with 
$\xi^{\mu}_{\scriptscriptstyle (h)} \: \nabla_{\mu} \Omega = 0$ are 
also the trivial transformations, as in the $\Omega = const.$ case. 

Unusual behavior of the surface gravity, particular to 
asymptotic Killing horizons, occurs when we consider 
the supertranslations which non-trivially depend on the null coordinate, 
i.e., $\xi^{\mu}_{\scriptscriptstyle (h)} \: \nabla_{\mu} \Omega \neq 0$. 
To see this explicitly, we again analyze the supertranslations 
dependent only on the null coordinate, 
which are described by $\omega(u)$. 
By decomposing $\omega(u)$ into the Fourier modes, 
as Eq.(\ref{eqn:SuperTransDecomp}), $\Omega$ for each mode 
is given as 
\begin{equation} 
\Omega_{k}(u) = i \: k \: \varpi_{k}
\left( \kappa_0 \: u \right)^{i k} \; , 
\label{eqn:OmegaDiffSExp} 
\end{equation} 
where we restored the infinitesimal coefficients $\varpi_{k}$. 
Then, the surface gravity $\kappa$ is transformed as 
\begin{equation}
\kappa  \longrightarrow \widetilde{\kappa} = 
\bigl[ 1 -  ( 1 + i \: k ) \: \Omega_{k}(u) \bigr] \: \kappa \; , 
\label{eqn:KappaTrans} 
\end{equation} 
while the horizon generating vector 
$\xi^{\mu}_{\scriptscriptstyle (h)}$ changes as 
\begin{equation} 
\xi^{\mu}_{\scriptscriptstyle (h)} \longrightarrow 
\widetilde{\xi}^{\mu}_{\scriptscriptstyle (h)} = 
\bigl[ 1 - \Omega_{k}(u) \bigr] \:   
\xi^{\mu}_{\scriptscriptstyle (h)} \; , 
\label{eqn:XiTrans} 
\end{equation}
on the horizon. 

As we see from Eqs.(\ref{eqn:KappaTrans}) and (\ref{eqn:XiTrans}), 
$\tilde{\kappa}$ and $\tilde{\xi}^{\mu}_{\scriptscriptstyle (h)}$ now 
oscillate along the null direction on the horizon, and we cannot 
re-normalize those variables into constant values over the horizon 
as in the previous argument, which already exhibits unusual behavior 
of asymptotic Killing horizons. 
Instead, it is sufficient, for the present purpose, 
to analyze the values of those variables 
on an arbitrary fixed horizon two-sphere, which is specified by 
$u = u_0 = const.$. 
When we adjust $\varpi_{k}$ in Eq.(\ref{eqn:OmegaDiffSExp}) 
so that $\Omega_{k}$ is real on the horizon 
two-sphere $u = u_0$, 
the horizon generating vector 
$\tilde{\xi}^{\mu}_{\scriptscriptstyle (h)}$ also remains real 
on $u = u_0$. 
However, $\tilde{\kappa}$ is complex even on that horizon two-sphere. 
Correspondingly, 
when we re-normalize $\tilde{\xi}^{\mu}_{\scriptscriptstyle (h)}$ 
by the factor $\left[ 1 - \Omega_{k}(u_0) \right]^{-1}$, 
it coincides with its original value 
$\xi^{\mu}_{\scriptscriptstyle (h)}$ on the horizon two-sphere 
$u = u_0$, but the surface gravity on $u = u_0$ changes as 
\begin{equation} 
\widetilde{\kappa} \longrightarrow 
\bigl[ 1 -  i \: k \: \Omega_{k}(u_0) \bigr] \: \kappa \; . 
\end{equation} 
Thus, the surface gravity cannot be brought back to the original value, 
due to the presence of the anomalous factor $- i \: k \: \Omega_{k}$, 
even when the horizon generating vector is re-normalized to 
the original value and the metric is fixed. 
This behavior of the surface gravity is clearly contrasted with 
the case of 
$\xi^{\mu}_{\scriptscriptstyle (h)} \: \nabla_{\mu} \Omega = 0$. 
We see that the supertranslations that depend non-trivially 
on the null coordinate result in the distinguishable feature of 
asymptotic Killing horizons  
from exact Killing horizons, while the local properties of 
asymptotic Killing horizons are preserved under those transformations. 

In fact, it is possible to re-normalize the horizon generating vector 
as $\tilde{\xi}^{\mu}_{\scriptscriptstyle (h)} \rightarrow 
\left[ 1 - ( 1 + i \: k ) \: \Omega_{k}(u_0)  \right]^{-1} \, 
\tilde{\xi}^{\mu}_{\scriptscriptstyle (h)}$, so that 
the surface gravity is re-normalized into its original value $\kappa$ 
on $u = u_0$. 
However, $\tilde{\xi}^{\mu}_{\scriptscriptstyle (h)}$ is now a 
complex-valued vector on $u = u_0$. 
Since the horizon generating vector 
before the transformation gives the directional derivative along 
the Killing time, we may speculate that 
the horizon deformations of 
$\xi^{\mu}_{\scriptscriptstyle (h)} \: \nabla_{\mu} \Omega \neq 0$ 
together with this re-normalization will describe 
analytic continuation of the Killing time into a complex plane. 
From this point of view, it will be interesting to explore 
in future investigations the possibility 
that the asymptotic symmetries are related to the Euclidean approach 
to black hole thermodynamics\footnote{
The Euclidean approach to black hole thermodynamics played a 
crucial role to choose a specific form of the symmetry vectors 
in Ref.\cite{Carlip1,Carlip2}.}.

\section{Poisson bracket algebra} 
\label{sec:PoissonBracket} 
\eqnum{0} 

In this section, we analyze the Poisson bracket algebra of 
conserved charges, by employing the covariant phase space formalism 
developed by Wald and 
his collaborators\cite{CovariantPhase,WaldZoupas}. 
After we summarize the definition of conserved charges, 
we consider the Poisson bracket algebra associated with 
the asymptotic symmetries on Killing horizons. 
Because we need to specify an explicit form of a Lagrangian in order to 
calculate the Poisson bracket algebra, 
now we focus on the Einstein theory with 
or without the cosmological constant. 
We also consider extension of the asymptotic symmetries and 
discuss whether and how 
the Poisson bracket algebra acquires a central charge. 

\subsection{Conserved charges in the covariant phase space} 

The covariant phase space is defined as the space of solutions to 
field equations with boundary conditions imposed. 
Then, conserved charges, such as a Hamiltonian and angular momenta, 
are defined through variations of a Lagrangian density. 
For the Lagrangian $L(\psi)$, where $\psi^{a}$ denotes 
the dynamical field variables collectively, 
variations of the Lagrangian density are expressed as 
\begin{equation}
\overline{\delta} \, \Bigl( \varepsilon_{\mu \nu \rho \sigma} \: L(\psi) 
\Bigr) = \varepsilon_{\mu \nu \rho \sigma} \, E_{a} \: 
\overline{\delta} \psi^{a} 
+ \varepsilon_{\mu \nu \rho \sigma} \,  
\nabla_{\beta} \, \Theta^{\beta}(\psi; \overline{\delta} \psi) \; , 
\label{eqn:ActionVariation}
\end{equation} 
when the dynamical field variables are varied as 
$\psi^{a} \rightarrow \psi^{a} + \overline{\delta} \psi^{a}$, 
where $E_{a} = 0$ are the field equations 
and $\Theta^{\beta}(\psi; \overline{\delta} \psi)$ in the totally divergent term depends linearly on $\overline{\delta} \psi^{a}$. 

Since variations $\overline{\delta}$ are arbitrary, 
we can consider that those are generated by the Lie derivatives 
along arbitrary vectors $\varrho^{\mu}$, as 
$\overline{\delta} \psi^{a} = {\cal L}_{\varrho} \psi^{a}$. 
Then, by defining the vector $J^{\beta}(\psi; \varrho)$ as 
$J^{\beta}(\psi; \varrho) \equiv 
\Theta^{\beta}(\psi; {\cal L}_{\varrho} \psi) - 
\varrho^{\beta} \, L(\psi)$, 
we see that $\nabla_{\beta} \, J^{\beta}(\psi; \varrho) = 0$ 
if the field equations $E_{a} = 0$ are satisfied. 
We can also find the antisymmetric tensor 
$Q^{\beta \alpha}(\psi; \varrho)$, such that its divergence gives 
$J^{\beta}(\psi; \varrho)$ as 
$J^{\beta}(\psi; \varrho) \equiv \nabla_{\alpha} \, 
Q^{\beta \alpha}(\psi; \varrho)$ 
when the field equations hold. 

Furthermore, we define the symplectic current density 
$\omega_{\mu \nu \rho}(\psi; \overline{\delta}_{1} \psi, 
\overline{\delta}_{2} \psi)$ as 
\begin{equation} 
\omega_{\mu \nu \rho}(\psi; \overline{\delta}_{1} \psi, 
\overline{\delta}_{2} \psi) 
\equiv 
\overline{\delta}_{2} \Bigl( \epsilon_{\beta \mu \nu \rho} \, 
\Theta^{\beta}(\psi; \overline{\delta}_{1} \psi) \Bigr) 
- \overline{\delta}_{1} \Bigl( \epsilon_{\beta \mu \nu \rho} \, 
\Theta^{\beta}(\psi; \overline{\delta}_{2} \psi) \Bigr) \; ,  
\label{eqn:omegaDef} 
\end{equation} 
for arbitrary variations $\overline{\delta}_{1}$ and 
$\overline{\delta}_{2}$. Then, we find that it is conserved, 
\begin{equation} 
d_{\lambda} \, \omega_{\mu \nu \rho}(\psi; \overline{\delta}_{1} \psi, 
\overline{\delta}_{2} \psi) = 0 \; , 
\label{eqn:LinearFieldEq} 
\end{equation}  
if the linearized field equations $\overline{\delta}_{1} E_{a} = 0$ and 
$\overline{\delta}_{2} E_{a} = 0$ are satisfied, 
where $d_{\lambda}$ denotes the exterior derivative. 
Particularly, by setting as 
$\overline{\delta}_{1} \psi^{a} = {\cal L}_{\varrho} \psi^{a}$, 
the conserved charges $\mbox{\large ${\cal H}$}[\psi; \varrho]$ 
conjugate to arbitrary vectors $\varrho^{\mu}$ 
on a partial Cauchy surface ${\cal C}$ are defined by 
\begin{equation} 
\overline{\delta} \, \mbox{\large ${\cal H}$}[\psi, \varrho] \equiv 
\int_{\cal C} \omega_{\mu \nu \rho}
(\psi; {\cal L}_{\varrho} \psi, \overline{\delta} \psi) \; . 
\label{eqn:DefHamiltonian} 
\end{equation} 
We emphasize here that the variations denoted by $\overline{\delta}$ 
act only on the dynamical field variables $\psi^{a}$, 
but not on the non-dynamical variable $\varrho^{\mu}$, 
and that the derivations of the key relations below, 
such as Eqs.(\ref{eqn:omegaTotDiv}) and (\ref{eqn:HamiltonianExist}), 
are based on this fact. 

We can show that when the linearized field equations are satisfied, 
$\omega_{\mu \nu \rho}(\psi; {\cal L}_{\varrho} \psi, 
\overline{\delta} \psi)$ is written as 
\begin{equation} 
\omega_{\mu \nu \rho}(\psi; {\cal L}_{\varrho} \psi, 
\overline{\delta} \psi) 
= d_{\mu} \biggl[ \overline{\delta} \Bigl( \frac{1}{2} \: 
\varepsilon_{\beta \alpha \nu \rho} \: Q^{\beta \alpha}(\psi; \varrho) 
\Bigr) 
+ \, \varepsilon_{\beta \alpha \nu \rho} \, \varrho^{\beta} \, 
\Theta^{\alpha}(\psi; \overline{\delta} \psi) \biggr] \; ,  
\label{eqn:omegaTotDiv} 
\end{equation} 
in the covariant phase space, 
and hence the conserved charges are expressed in the form of  
an integral over the boundary $\partial {\cal C}$ of ${\cal C}$, as 
\begin{equation} 
\overline{\delta} \, \mbox{\large ${\cal H}$}[\psi; \varrho] 
= \overline{\delta} \int_{\partial {\cal C}} \frac{1}{2} \: 
\varepsilon_{\beta \alpha \mu \nu} \: Q^{\beta \alpha}(\psi; \varrho)  
+ \int_{\partial {\cal C}} \varepsilon_{\beta \alpha \mu \nu} \, 
\varrho^{\beta} \, \Theta^{\alpha}(\psi; \overline{\delta} \psi) \; . 
\label{eqn:HamiltonianSurface} 
\end{equation} 
As we can see from Eq.(\ref{eqn:HamiltonianSurface}), however, 
$\overline{\delta} \, \mbox{\large ${\cal H}$}[\psi; \varrho]$ is not 
defined as a total variation, and then integrability of 
Eq.(\ref{eqn:HamiltonianSurface}) is not guaranteed generally. 
It is necessary that two arbitrary variations tangent to the covariant 
phase space, 
$\overline{\delta}_{1}$ and $\overline{\delta}_{2}$, 
should commute when they act on 
$\mbox{\large ${\cal H}$}[\psi; \varrho]$, i.e., 
$( \overline{\delta}_{1} \overline{\delta}_{2} 
- \overline{\delta}_{2} \overline{\delta}_{1} ) 
\mbox{\large ${\cal H}$}[\psi; \varrho] = 0$, in order that 
Eq.(\ref{eqn:HamiltonianSurface}) is integrable and the well-defined 
conserved charges $\mbox{\large ${\cal H}$}[\psi; \varrho]$ exist. 
Indeed, Wald and Zoupas\cite{WaldZoupas} showed that it is 
the necessary and sufficient condition 
for the existence of $\mbox{\large ${\cal H}$}[\psi; \varrho]$, 
which is rewritten as  
\begin{equation} 
\int_{\partial {\cal C}} \varrho^{\beta} \, 
\omega_{\beta \mu \nu}(\psi; \overline{\delta}_{1} \psi, 
\overline{\delta}_{2} \psi) = 0 \; . 
\label{eqn:HamiltonianExist} 
\end{equation} 
If the well-defined conserved charges 
$\mbox{\large ${\cal H}$}[\psi; \varrho]$ exist, 
the second term of the right-hand side of 
Eq.(\ref{eqn:HamiltonianSurface}) must be written as a total variation, 
and hence there exists a vector $B^{\alpha}(\psi)$, such that 
\begin{equation} 
\overline{\delta} \int_{\partial {\cal C}} 
\varepsilon_{\beta \alpha \mu \nu} \, 
\varrho^{\beta} \, B^{\alpha}(\psi) \equiv
\int_{\partial {\cal C}} \varepsilon_{\beta \alpha \mu \nu} \,  
\varrho^{\beta} \, \Theta^{\alpha}(\psi; \overline{\delta} \psi) 
\; . 
\label{eqn:BDef} 
\end{equation} 
We then have 
\begin{equation} 
\mbox{\large ${\cal H}$}[\psi; \varrho] = 
\int_{\partial {\cal C}} \frac{1}{2} \: 
\varepsilon_{\beta \alpha \mu \nu} \Bigl[ 
Q^{\beta \alpha}(\psi; \varrho) 
+ 2 \: \varrho^{[ \beta} \, B^{\alpha ]}(\psi) \Bigr] 
+ \mbox{\large ${\cal H}$}_{0}[\varrho] 
\; , 
\label{eqn:HamiltonianInt} 
\end{equation} 
where $\mbox{\large ${\cal H}$}_{0}[\varrho]$ is an integration 
constant in the sense 
that it does not depend on the dynamical field variables.

\subsection{Poisson bracket algebra of the asymptotic symmetries} 

Now we consider the Poisson bracket algebra of 
the conserved charges $\mbox{\large ${\cal H}$}[g; \zeta]$ conjugate to 
asymptotic Killing vectors $\zeta^{\mu}$. Here, we assume that 
the background spacetime is described by the Einstein theory with or 
without the cosmological constant, where the Lagrangian is given by 
\begin{equation} 
L(g) \equiv \frac{1}{16 \pi} \, ( R - 2 \Lambda ) \; . 
\label{eqn:Lagrangian} 
\end{equation} 

In order to define $\mbox{\large ${\cal H}$}[g; \zeta]$, we have to 
specify a partial Cauchy surface ${\cal C}$ in 
Eq.(\ref{eqn:DefHamiltonian}), or equivalently, its boundary 
$\partial {\cal C}$ in Eq.(\ref{eqn:HamiltonianSurface}). 
Since we have imposed the boundary condition 
Eq.(\ref{eqn:LimitMetricDS}) on the null hypersurface $v = 0$, 
it will be appropriate to consider that 
a two-sphere on $v = 0$ comprises a part of $\partial {\cal C}$. 
Here, Cauchy surfaces can arbitrarily intersect that null hypersurface, 
so that they need not be degenerate at $v = 0$, 
and two-spheres at $v = 0$ are not necessarily the bifurcation surface of 
the Killing horizon in the background spacetime. 
We should also take into account other possible parts of 
$\partial {\cal C}$, if any, such as a large two-sphere at infinity.  
However, we can adjust the form of $\zeta^{\mu}$, 
such that those parts of $\partial {\cal C}$, 
other than a two-sphere at $v = 0$, 
do not make any contribution to the conserved charges. 
This will be justified because 
there exist no equations that govern the behavior of $\zeta^{\mu}$ 
throughout the spacetime. 
Therefore, here we consider only the contribution from 
a two-sphere at $v = 0$. 

We must also verify that $\mbox{\large ${\cal H}$}[g; \zeta]$ are 
well-defined under the boundary condition Eq.(\ref{eqn:LimitMetricDS}) 
at $v = 0$. We thus examine whether Eq.(\ref{eqn:HamiltonianExist}) 
is satisfied for $\overline{\delta} g_{\mu \nu} = {\cal O}(v)$, 
by calculating 
$\omega_{\mu \nu \rho}(g; \overline{\delta}_{1} g, 
\overline{\delta}_{2} g)$ for the 
Lagrangian Eq.(\ref{eqn:Lagrangian}), where 
$\Theta^{\beta}(g; \overline{\delta} g)$ is given\footnote{
$\Theta^{\beta}(g; \overline{\delta} g)$ in 
Eq.(\ref{eqn:ThetaEinstein}) and $Q^{\beta \alpha}(g; \zeta)$ in 
Eq.(\ref{eqn:QEinstein}) are defined only up to 
ambiguity\cite{CovariantPhase}. 
Here, we make a natural choice, and effects of the ambiguity 
will be discussed in future investigations.} by  
\begin{equation} 
\Theta^{\beta}(g; \overline{\delta} g) = \frac{1}{16 \pi} 
\Bigl[ g_{\mu \nu} \, \nabla^{\beta} \, \overline{\delta} \, g^{\mu \nu} 
- \nabla_{\nu} \, \overline{\delta} \, g^{\nu \beta} \Bigr] \; . 
\label{eqn:ThetaEinstein} 
\end{equation} 
By noticing that $\overline{\delta}_{2} \, 
\Theta^{\beta}(g; \overline{\delta}_{1} g) 
- \overline{\delta}_{1} \, 
\Theta^{\beta}(g; \overline{\delta}_{2} g)$ is proportional to either 
$\overline{\delta}_1 g_{\mu \nu}$ or   
$\overline{\delta}_2 g_{\mu \nu}$, we obtain 
\begin{eqnarray} 
\! & \! \! \! \! \! \! & \! 
\omega_{\mu \nu \rho}(g; \overline{\delta}_{1} g, 
\overline{\delta}_{2} g) \nonumber \\ 
\! & \! \! \! = \! \! \! & \! 
\Bigl( \, \overline{\delta}_{2} \varepsilon_{\beta \mu \nu \rho} 
\Bigr) \,  
\Theta^{\beta}(g; \overline{\delta}_{1} g) 
- \Bigl( \, \overline{\delta}_{1} \varepsilon_{\beta \mu \nu \rho} 
\Bigr) \, 
\Theta^{\beta}(g; \overline{\delta}_{2} g) 
+ \; \varepsilon_{\beta \mu \nu \rho} \, 
\Bigl( \, \overline{\delta}_{2} \, 
\Theta^{\beta}(g; \overline{\delta}_{1} g) 
- \overline{\delta}_{1} \, 
\Theta^{\beta}(g; \overline{\delta}_{2} g) \Bigr) 
\nonumber \\ 
\! & \! \! \! = \! \! \! & \! {\cal O}(v) \; . 
\end{eqnarray} 
Since $\omega_{\mu \nu \rho}(g; \overline{\delta}_{1} g, 
\overline{\delta}_{2} g)$ 
vanishes at $v = 0$, Eq.(\ref{eqn:HamiltonianExist}) is satisfied, 
and hence we see that the conserved charges 
$\mbox{\large ${\cal H}$} [g; \zeta]$ are well-defined  
under the boundary condition Eq.(\ref{eqn:LimitMetricDS}). 

Once the conserved charges are found to be well-defined, 
we can consider the Poisson brackets 
between those conserved charges, which will be written as 
\begin{equation} 
\Bigl\{ \mbox{\large ${\cal H}$}[g; \zeta_{1}] 
\; , \; \mbox{\large ${\cal H}$}[g; \zeta_{2}] \Bigr\} 
=  \overline{\delta}_{\zeta_{2}} 
\mbox{\large ${\cal H}$}[g; \zeta_{1}] \; , 
\label{eqn:PoissonBra1} 
\end{equation} 
in the covariant phase space. Here, $\overline{\delta}_{\zeta_{2}}$ 
denotes the variation generated by the Lie derivative of the metric 
along $\zeta^{\mu}_{2}$, whose action is thus defined by  
$\overline{\delta}_{\zeta_{2}} g_{\mu \nu} 
\equiv {\cal L}_{\zeta_{2}} g_{\mu \nu}$ and 
$\overline{\delta}_{\zeta_{2}} \zeta^{\mu}_{1} \equiv 0$. 
In the context of the previous section, 
$\overline{\delta}_{\zeta_{2}}$ stands for nothing but 
the dynamical field transformation generated by the 
asymptotic Killing vector $\zeta_{2}^{\mu}$. 

If the Poisson bracket algebra is isomorphic to the Lie bracket algebra, 
Eq.(\ref{eqn:PoissonBra1}) will be equal to  
$\mbox{\large ${\cal H}$}[g; {\cal L}_{\zeta_{1}} \zeta_{2}]$. 
However, as in the case of the asymptotic symmetries in 
the 3-d anti-de Sitter spacetime\cite{BrownHenneaux}, 
this is not the case in general (see also Ref.\cite{CovariantPhase} 
for the relevant issue). 
By recalling that $\overline{\delta}$ and 
$\overline{\delta}_{\zeta_{2}}$ commute when they act on the 
well-defined conserved charges, 
and noticing that ${\cal L}_{\zeta_{2}} \overline{\delta} 
\mbox{\large ${\cal H}$}[g; \zeta_{1}] = \overline{\delta}_{\zeta_{2}} 
\overline{\delta} \mbox{\large ${\cal H}$}[g; \zeta_{1}] + 
\overline{\delta} 
\mbox{\large ${\cal H}$}[g; {\cal L}_{\zeta_{2}} \zeta_{1}]$, 
we actually have 
\begin{eqnarray} 
\overline{\delta} \, \Bigl( \, \overline{\delta}_{\zeta_{2}} 
\mbox{\large ${\cal H}$}[g; \zeta_{1}] - 
\mbox{\large ${\cal H}$}[g; {\cal L}_{\zeta_{1}} \zeta_{2}] \Bigr) 
\! & \! \! \! = \! \! \! & \! 
{\cal L}_{\zeta_{2}} \,  \overline{\delta} \, 
\mbox{\large ${\cal H}$}[g; \zeta_{1}] 
\nonumber \\ 
\! & \! \! \! = \! \! \! & \! 
\int_{\cal C} {\cal L}_{\zeta_{2}} \, 
\omega_{\mu \nu \rho}(g; {\cal L}_{\zeta_{1}} g, \overline{\delta} g) 
\nonumber \\ 
\! & \! \! \! = \! \! \! & \! 
\int_{\cal C} \Bigl[ \zeta^{\beta}_{2} \, d_{\beta} \, 
\omega_{\mu \nu \rho}(g; {\cal L}_{\zeta_{1}} g, \overline{\delta} g) 
+ d_{\mu} \bigl( \zeta^{\beta}_{2} \, 
\omega_{\beta \nu \rho}(g; {\cal L}_{\zeta_{1}} g, \overline{\delta} g) 
\bigr) \Bigr] \; . 
\label{eqn:VariationCentral} 
\end{eqnarray} 
The right-hand side of Eq.(\ref{eqn:VariationCentral}) vanishes, 
if Eqs.(\ref{eqn:LinearFieldEq}) and (\ref{eqn:HamiltonianExist}) 
are satisfied. As we have seen above, Eq.(\ref{eqn:HamiltonianExist}) 
is indeed satisfied. In addition, Eq.(\ref{eqn:LinearFieldEq}) also holds 
when the variations $\overline{\delta}$ are tangent to the covariant 
phase space, so that the perturbed metrics 
$\overline{\delta} g_{\mu \nu}$ satisfy 
the linearized Einstein equation. 
Here, we consider only those variations, as we did when we derived 
Eqs.(\ref{eqn:omegaTotDiv}) and (\ref{eqn:HamiltonianExist}), 
and integrate Eq.(\ref{eqn:VariationCentral}) 
in the covariant phase space. 
We then obtain 
\begin{equation} 
\overline{\delta}_{\zeta_{2}} 
\mbox{\large ${\cal H}$}[g; \zeta_{1}] 
= \mbox{\large ${\cal H}$}[g; {\cal L}_{\zeta_{1}} \zeta_{2}] 
+ \mbox{\large $K$}[\zeta_{1}, \zeta_{2}]  
\; , 
\label{eqn:TruePoisson} 
\end{equation} 
where $\mbox{\large $K$}[\zeta_{1}, \zeta_{2}]$ is constant 
under the variations, i.e., 
$\overline{\delta} \mbox{\large $K$}[\zeta_{1}, \zeta_{2}] = 0$, 
and is referred to as the central term. 
Therefore, from Eqs.(\ref{eqn:PoissonBra1}) and 
(\ref{eqn:TruePoisson}), we find that 
the Poisson bracket algebra of the conserved charges 
$\mbox{\large ${\cal H}$}[g; \zeta]$ is expressed as 
\begin{equation} 
\Bigl\{ \mbox{\large ${\cal H}$}[g; \zeta_{1}] \; , \; 
\mbox{\large ${\cal H}$}[g; \zeta_{2}] \Bigr\} = 
\mbox{\large ${\cal H}$}[g; {\cal L}_{\zeta_{1}} \zeta_{2}] 
+ \mbox{\large $K$}[\zeta_{1}, \zeta_{2}]  \; . 
\end{equation} 

We should evaluate the central term 
$\mbox{\large $K$}[\zeta_{1}, \zeta_{2}]$ to 
establish the Poisson bracket algebra. 
Since the central term $\mbox{\large $K$}[\zeta_{1}, \zeta_{2}]$ 
does not depend on the dynamical field variables, 
we can calculate it on the background spacetime ${}^{0}\!g_{\mu \nu}$, 
as 
\begin{equation} 
\mbox{\large $K$}[\zeta_{1}, \zeta_{2}] = \overline{\delta}_{\zeta_{2}} 
\mbox{\large ${\cal H}$}[{}^{0}\!g; \zeta_{1}] 
- \mbox{\large ${\cal H}$}[{}^{0}\!g; {\cal L}_{\zeta_{1}} \zeta_{2}] 
= {\cal L}_{\zeta_2} 
\mbox{\large ${\cal H}$}[{}^{0}\!g; \zeta_{1}] \; . 
\label{eqn:EvaluateK} 
\end{equation}  
On the other hand, by substituting the definitions of 
$B^{\alpha}(g)$, $J^{\alpha}(g; \zeta)$ 
and $Q^{\alpha \beta}(g; \zeta)$ into the Lie derivative of 
Eq.(\ref{eqn:HamiltonianInt}), 
and integrating by parts, we obtain 
\begin{eqnarray} 
{\cal L}_{\zeta_2} 
\mbox{\large ${\cal H}$}[{}^{0}\!g; \zeta_{1}] 
\! & \! \! \! = \! \! \! & \! 
- \; \mbox{\large ${\cal H}$}[{}^{0}\!g; 
{\cal L}_{\zeta_{1}} \zeta_{2}]
+ \int_{\partial {\cal C}} \frac{3}{2} \: 
{}^{0}\!\varepsilon_{\beta \alpha \mu \nu} \, 
{}^{0} \nabla_{\gamma} \Bigl( \zeta^{[ \beta}_{1} \, 
Q^{\alpha \gamma ]}({}^{0}\!g; \zeta_{2}) \Bigr) \nonumber \\ 
& & \! + \int_{\partial {\cal C}} \frac{1}{2} \: 
{}^{0}\!\varepsilon_{\beta \alpha \mu \nu} \Bigl[ {\cal L}_{\zeta_{2}} 
Q^{\beta \alpha}({}^{0}\!g; \zeta_{1}) - {\cal L}_{\zeta_{1}} 
Q^{\beta \alpha}({}^{0}\!g; \zeta_{2}) 
- Q^{\beta \alpha}({}^{0}\!g; {\cal L}_{\zeta_{2}} \zeta_{1}) 
\nonumber \\ 
& & \! 
+ \; Q^{\beta \alpha}({}^{0}\!g; \zeta_{1}) \, 
\bigl( {}^{0} \nabla_{\gamma} \, \zeta^{\gamma}_{2} \bigr) 
- Q^{\beta \alpha}({}^{0}\!g; \zeta_{2}) \, 
\bigl( {}^{0} \nabla_{\gamma} \, \zeta^{\gamma}_{1} \bigr) 
+ 2 \: \zeta^{[ \beta}_{1} \, \zeta^{\alpha ]}_{2} \, L({}^{0}\!g) \Bigr] 
\; . 
\label{eqn:CentralEvalpre1} 
\end{eqnarray} 
The second term in the right-hand side of Eq.(\ref{eqn:CentralEvalpre1}) 
does not contribute, because it is an integral of a total divergence 
on the two-sphere $\partial {\cal C}$, which has no boundary. 
Then, Eqs.(\ref{eqn:EvaluateK}) and (\ref{eqn:CentralEvalpre1}) are 
described\footnote{ 
Eq.(\ref{eqn:CentralChargeEin}) does not coincide 
with the corresponding expression in Ref.\cite{Carlip2}. 
This is because the variation corresponding to 
$\overline{\delta}_{\zeta_{2}}$ acts on the conserved charges 
as the Lie derivative ${\cal L}_{\zeta_{2}}$ in Ref.\cite{Carlip2}.} as 
\begin{eqnarray} 
\overline{\delta}_{\zeta_{2}} 
\mbox{\large ${\cal H}$}[{}^{0}\!g; \zeta_{1}] 
\! & \! \! \! = \! \! \! & \! 
\int_{\partial {\cal C}} \frac{1}{2} \: 
{}^{0}\!\varepsilon_{\beta \alpha \mu \nu} \Bigl[ {\cal L}_{\zeta_{2}} 
Q^{\beta \alpha}({}^{0}\!g; \zeta_{1}) - {\cal L}_{\zeta_{1}} 
Q^{\beta \alpha}({}^{0}\!g; \zeta_{2}) 
- Q^{\beta \alpha}({}^{0}\!g; {\cal L}_{\zeta_{2}} \zeta_{1}) 
\nonumber \\ 
& & \! 
+ \; Q^{\beta \alpha}({}^{0}\!g; \zeta_{1}) \, 
\bigl( {}^{0} \nabla_{\gamma} \, \zeta^{\gamma}_{2} \bigr) 
- Q^{\beta \alpha}({}^{0}\!g; \zeta_{2}) \, 
\bigl( {}^{0} \nabla_{\gamma} \, \zeta^{\gamma}_{1} \bigr) 
+ 2 \: \zeta^{[ \beta}_{1} \, \zeta^{\alpha ]}_{2} \, L({}^{0}\!g) \Bigr] 
\nonumber \\ 
\! & \! \! \! = \! \! \! & \! 
\int_{\partial {\cal C}} \frac{1}{16 \pi} 
{}^{0}\!\varepsilon_{\beta \alpha \mu \nu} \: 
\Bigl[ \bigl( {}^{0} \nabla^{\gamma} \zeta^{\alpha}_{1} \bigr) 
 \bigl( {}^{0} \nabla_{\gamma} \zeta^{\beta}_{2} \bigr) 
- \bigl( {}^{0} \nabla^{\gamma} \zeta^{\beta}_{1} \bigr) 
 \bigl( {}^{0} \nabla_{\gamma} \zeta^{\alpha}_{2} \bigr) 
- {}^{0}\!R^{\beta \alpha}_{~~~ \gamma \lambda} \: \zeta^{\gamma}_{1} 
\zeta^{\lambda}_{2} 
\nonumber \\ 
\! & \! \! \! + \! \! \! & \! \! \! 
\zeta^{[\beta}_{1} \zeta^{\alpha]}_{2} 
\bigl( {}^{0}\!R - 2 \Lambda \bigr) 
+ \frac{1}{2} \bigl( {}^{0} \nabla^{\alpha} \zeta^{\beta}_{1} 
-  {}^{0} \nabla^{\beta} \zeta^{\alpha}_{1} \bigr) 
\bigl( {}^{0} \nabla_{\gamma} \zeta^{\gamma}_{2} \bigr) 
- \frac{1}{2} \bigl( {}^{0} \nabla^{\alpha} \zeta^{\beta}_{2} 
-  {}^{0} \nabla^{\beta} \zeta^{\alpha}_{2} \bigr) 
\bigl( {}^{0} \nabla_{\gamma} \zeta^{\gamma}_{1} \bigr) \Bigr] \, ,   
\label{eqn:CentralChargeEin}
\end{eqnarray} 
where we substituted the form of $Q^{\beta \alpha}(g; \zeta)$ 
for the Lagrangian Eq.(\ref{eqn:Lagrangian}), which is given by 
\begin{equation} 
Q^{\beta \alpha}(g; \zeta) = \frac{1}{16 \pi} \Bigl[ \nabla^{\alpha} 
\zeta^{\beta} - \nabla^{\beta} \zeta^{\alpha} \Bigr] \; .  
\label{eqn:QEinstein} 
\end{equation} 
We note that the derivation of Eq.(\ref{eqn:CentralChargeEin}) 
is not based on the specific properties to 
asymptotic Killing vectors $\zeta^{\mu}$, 
and hence it is valid for arbitrary vectors. 
In particular, we can reproduce, 
by using Eq.(\ref{eqn:CentralChargeEin}), the correct value of 
the central charge of the Poisson bracket algebra associated with 
the asymptotic symmetries in the 
3-d anti-de Sitter spacetime\cite{BrownHenneaux}. 

As we describe in Appendix \ref{sec:CalcCenter}, 
when we substitute Eq.(\ref{eqn:ZetaCom}), the Einstein equation and 
the exact Killing equations for 
$\xi^{\mu}_{\scriptscriptstyle (h)}$ and 
$\xi^{\mu}_{\scriptscriptstyle (i)}$ in the 
background spacetime, Eq.(\ref{eqn:CentralChargeEin}) 
reduces to 
\begin{equation} 
\overline{\delta}_{\zeta_{2}} 
\mbox{\large ${\cal H}$}[{}^{0}\!g; \zeta_{1}] = 0 
\; , 
\label{eqn:VanishingC} 
\end{equation} 
where we used regularity of asymptotic Killing vectors, 
as well as the facts that 
$\xi^{\mu}_{\scriptscriptstyle (h)}$ is hypersurface orthogonal 
and proportional to $\nabla^{\mu} v$. 
Therefore, we have 
$\mbox{\large $K$}[\zeta_{1}, \zeta_{2}] 
= - \: \mbox{\large ${\cal H}$}[{}^{0}\!g; {\cal L}_{\zeta_{1}} \zeta_{2}]$, 
and the Poisson bracket algebra of the conserved charges is given by 
\begin{equation} 
\Bigl\{ \mbox{\large ${\cal H}$}[g; \zeta_{1}] \; , \; 
\mbox{\large ${\cal H}$}[g; \zeta_{2}] \Bigr\} = 
\mbox{\large ${\cal H}$}[g; {\cal L}_{\zeta_{1}} \zeta_{2}]
- \mbox{\large ${\cal H}$}[{}^{0}\!g; {\cal L}_{\zeta_{1}} \zeta_{2}] 
\; . 
\label{eqn:FaulseCentral} 
\end{equation} 
The last term of the right-hand side of Eq.(\ref{eqn:FaulseCentral}) 
looks like a central charge. 
However, we should examine whether it can be eliminated 
by constant shifts of the values of the conserved charges, 
or redefinition of $\mbox{\large ${\cal H}$}_{0}[\zeta]$ in 
Eq.(\ref{eqn:HamiltonianInt}). 
Although the left-hand side of Eq.(\ref{eqn:FaulseCentral}) does not 
change under such shifts, the right-hand side does. 
In fact, if we redefine as 
\begin{equation} 
\mbox{\large ${\cal H}$}[g; \zeta] \rightarrow 
\mbox{\large ${\cal H}'$}[g; \zeta] \equiv 
\mbox{\large ${\cal H}$}[g; \zeta]
- \mbox{\large ${\cal H}$}[{}^{0}\!g; \zeta] \; , 
\label{eqn:RenormHamiltonian} 
\end{equation} 
such that $\mbox{\large ${\cal H}'$}[{}^{0}\!g; \zeta] = 0$, 
the Poisson bracket algebra becomes 
\begin{equation} 
\Bigl\{ \mbox{\large ${\cal H}'$}[g; \zeta_{1}] \; , \; 
\mbox{\large ${\cal H}'$}[g; \zeta_{2}] \Bigr\} 
= \mbox{\large ${\cal H}'$}[g; {\cal L}_{\zeta_{1}} \zeta_{2}] \; , 
\label{eqn:TrueCentral} 
\end{equation} 
where the central term has been eliminated, in contrast to 
the Virasoro algebra Eq.(\ref{eqn:Virasoro}) in the case of 
the 3-d anti-de Sitter spacetime\cite{BrownHenneaux}. 
Therefore, we see that the Poisson bracket algebra associated with 
the asymptotic symmetries on Killing horizons does not acquire 
non-trivial central charges. 
Particularly, the supertranslations dependent only on 
the null coordinate, defined by Eq.(\ref{eqn:SuperTransNull}), 
yield the Poisson bracket subalgebra 
\begin{equation} 
\Bigl\{ \mbox{\large ${\cal H}'$}[g; \zeta_{k}] \; , \; 
\mbox{\large ${\cal H}'$}[g; \zeta_{k'}] \Bigr\} 
= i \, ( k' - k ) \, \mbox{\large ${\cal H}'$}[g; \zeta_{k+k'}] \; , 
\end{equation} 
which is isomorphic to the $\mathit{Diff}(S^1)$ algebra.

\subsection{Extended symmetries} 

The Poisson bracket algebra without a central charge 
Eq.(\ref{eqn:TrueCentral}) resulted from the boundary condition 
$\overline{\delta} g_{\mu \nu} = {\cal O}(v)$. 
However, it may be possible that we obtain a central charge, 
if we consider a larger symmetry group by 
imposing a weaker boundary condition. 
While the asymptotic symmetry group we have analyzed so far 
preserves the translational invariance along the null direction 
as well as the spherical symmetries of a two-sphere at $v = 0$, 
event horizons in dynamically evolving spacetimes will not possess 
those symmetries. 
It may be necessary to take into account deformations of 
event horizons which violate those symmetries, in order to obtain 
the Virasoro algebra with non-trivial central charges. 
Then, it will be worth while considering extended symmetries by 
weakening the boundary condition. 

Although there will be many possible ways to weaken the boundary 
condition, essential structures of event horizons are required 
to be preserved. 
Since event horizons are generated by null geodesics in general, 
it will be reasonable to condition that null vectors 
on the Killing horizon in the background spacetime remain null 
even in perturbed spacetimes. 
Hence, we now impose the boundary condition $g_{u u} = {\cal O}(v)$ 
near the hypersurface $v = 0$. 
By assuming as before that this condition is preserved when 
perturbations of the metric are generated by 
the Lie derivatives along 
{\it extended symmetry vectors} $\varsigma^{\mu}$, we have 
\begin{equation} 
{\cal L}_{\varsigma} g_{u u} = {\cal O}(v) \; . 
\label{eqn:ExtendAsymKillEq} 
\end{equation} 
When we write $g_{\mu \nu}$ and $\varsigma^{\mu}$ as 
\begin{eqnarray} 
g_{\mu \nu} \! & \! \! \! = \! \! \! & \! 
\gamma_{\mu \nu}(u,\theta, \phi) + v \; h_{\mu \nu}(u, \theta, \phi) 
+ {\cal O}(v^2) \; , \\ 
\varsigma^{\mu} \! & \! \! \! = \! \! \! & \! 
\varsigma^{\mu}_{\scriptscriptstyle (0)} 
+ v \; \varsigma^{\mu}_{\scriptscriptstyle (1)} + {\cal O}(v^2) \; , 
\end{eqnarray} 
with $\gamma_{u u} = 0$, and require that 
Eq.(\ref{eqn:ExtendAsymKillEq}) holds 
for arbitrary perturbations $\gamma_{\mu \nu}$ and $h_{\mu \nu}$, 
we obtain $\varsigma^{v}_{\scriptscriptstyle (0)} = 0$, 
$\partial_{u} \, \varsigma^{\theta}_{\scriptscriptstyle (0)} = 0$ and 
$\partial_{u} \, \varsigma^{\phi}_{\scriptscriptstyle (0)} = 0$. 
Therefore, extended symmetry vectors $\varsigma^{\mu}$ 
are given as 
\begin{equation} 
\varsigma^{\mu} = \omega(u, \theta, \phi) 
\: \xi_{\scriptscriptstyle (h)}^{\mu} 
+ A^{\scriptscriptstyle (i)}(\theta, \phi) 
\: \xi^{\mu}_{\scriptscriptstyle (i)} 
+ v \; Z^{\mu} + {\cal O}(v^2) \; ,  
\label{eqn:ExtdAsymVec}
\end{equation} 
where $Z^{\mu}$ is an arbitrary vector, and 
$A^{\scriptscriptstyle (i)}(\theta, \phi)$ are arbitrary functions 
of the angular coordinates $\theta$ and $\phi$, 
only two of which are independent. 
Eq.(\ref{eqn:ExtdAsymVec}) is 
the general form of tangent vectors to the hypersurface $v = 0$, 
except that $\xi_{\scriptscriptstyle (h)}^{\mu} \nabla_{\mu} 
A^{\scriptscriptstyle (i)}(\theta, \phi) = 0$, which is required for the 
null vectors to remain null. 
Compared with asymptotic Killing vectors Eq.(\ref{eqn:ZetaCom}), 
$A^{\scriptscriptstyle (i)}(\theta, \phi)$ need not be constant 
and $Z^{\mu}$ can differ from $X^{\mu}(\omega)$ in the case of 
extended symmetry vectors.  

Correspondingly, while the Killing horizon in the background spacetime 
is a null hypersurface even 
in the perturbed spacetimes $g_{\mu \nu} = {}^{0}\!g_{\mu \nu} 
+ {\cal L}_{\varsigma} {}^{0}\!g_{\mu \nu}$, 
the group of the extended symmetries described by $\varsigma^{\mu}$ 
does not preserve the properties of asymptotic Killing horizons. 
The Killing horizon in the background spacetime may be deformed 
into a null hypersurface that does not look like a Killing horizon at all. 
Nevertheless, it will be meaningful to consider 
the extended symmetries in order to understand whether and how 
the Poisson bracket algebra acquires a central charge. 
Particularly, the conserved charges in the case of 
$\omega = \omega(\theta, \phi)$ have been 
analyzed in Ref.\cite{HottaEtAl}, where it is claimed that 
the algebra of the conserved charges does not involve any anomalous 
terms like a central charge. 

We then calculate the central term 
$\mbox{\large $K$}[\varsigma_{1}, \varsigma_{2}]$ 
in the Poisson bracket algebra of conserved charges conjugate to 
extended symmetry vectors $\varsigma^{\mu}$, 
by assuming that those conserved charges are well-defined. 
We can apply Eq.(\ref{eqn:CentralChargeEin}) by replacing 
$\zeta^{\mu}$ with $\varsigma^{\mu}$, because it is valid for arbitrary 
vectors, as we described before. 
From the derivation given in Appendix \ref{sec:CalcCenter}, we obtain 
\begin{equation} 
\mbox{\large $K$}[\varsigma_{1}, \varsigma_{2}] =
\int_{\partial {\cal C}} \frac{1}{16 \pi} 
{}^{0}\!\varepsilon_{\mu \nu} \Bigl[ 
\bigl( \xi_{\scriptscriptstyle (i)}^{\alpha} 
{}^{0} \nabla_{\alpha} A^{\scriptscriptstyle (i)}_1 \bigr) \, 
\chi_2 
- \bigl( \xi_{\scriptscriptstyle (i)}^{\alpha} 
{}^{0}\nabla_{\alpha} A^{\scriptscriptstyle (i)}_2  \bigr) \, 
\chi_1 \Bigr] 
- \mbox{\large ${\cal H}$}[{}^{0}\!g; 
{\cal L}_{\varsigma_{1}} \varsigma_{2}] 
\; , 
\label{eqn:CenterExtd} 
\end{equation} 
where ${}^{0}\!\varepsilon_{\mu \nu}$ is the volume element of 
a horizon two-sphere in the background spacetime 
and $\chi$ is defined by  
\begin{equation} 
\chi \equiv 
\xi_{\scriptscriptstyle (h)}^{\lambda} {}^{0}\nabla_{\lambda} \omega 
+ 2 \; \kappa_0 \; \omega - Z^{\lambda} \: {}^{0}\nabla_{\lambda} v \; .  
\label{eqn:ChiDef} 
\end{equation} 
We see that a non-trivial central charge does not appear 
unless $\xi_{\scriptscriptstyle (i)}^{\alpha} {}^{0}\nabla_{\alpha} 
A^{\scriptscriptstyle (i)} \neq 0$. 
Since 
$\xi_{\scriptscriptstyle (i)}^{\alpha} {}^{0}\nabla^{\beta} 
A^{\scriptscriptstyle (i)}$ gives rise to perturbations of 
the induced metric on a horizon two-sphere, it implies that we need to 
incorporate the diffeomorphism group of a spherically symmetric 
two-sphere in order to obtain a central charge. 

We have assumed above that conserved charges are well-defined. 
If well-defined conserved charges conjugate to 
extended symmetry vectors $\varsigma^{\mu}$ exist, however, 
it is necessary that Eq.(\ref{eqn:HamiltonianExist}) with 
$\psi^{a} = g_{\mu \nu}$ and $\varrho^{\mu} = \varsigma^{\mu}$ 
holds for $\overline{\delta} g_{\mu \nu} = 
{\cal L}_{\varsigma} {}^{0}\!g_{\mu \nu}$ 
on the background spacetime ${}^{0}\!g_{\mu \nu}$. 
Since $\xi_{\scriptscriptstyle (h)}^{\mu}$ belongs to 
extended symmetry vectors $\varsigma^{\mu}$, 
we particularly require 
\begin{equation} 
\int_{\partial {\cal C}} \xi_{\scriptscriptstyle (h)}^{\beta} \, 
\omega_{\beta \mu \nu}({}^{0}\!g; {\cal L}_{\varsigma_1} {}^{0}\!g, 
{\cal L}_{\varsigma_2} {}^{0}\!g) = 0 \; .  
\label{eqn:ExtndExistCharge} 
\end{equation} 
By substituting Eq.(\ref{eqn:ExtdAsymVec}), 
Eq.(\ref{eqn:ExtndExistCharge}) is rewritten as 
\begin{equation} 
\int_{\partial {\cal C}} {}^{0}\!\varepsilon_{\mu \nu} \Bigl[ 
\bigl( \xi_{\scriptscriptstyle (j)}^{\lambda} {}^{0}\nabla_{\lambda} 
A^{\scriptscriptstyle (j)}_2 \bigr) 
\bigl( \xi_{\scriptscriptstyle (h)}^{\alpha} {}^{0}\nabla_{\alpha} 
\chi_1 \bigr) 
- \bigl( \xi_{\scriptscriptstyle (j)}^{\lambda} {}^{0}\nabla_{\lambda} 
A^{\scriptscriptstyle (j)}_1 \bigr) 
\bigl( \xi_{\scriptscriptstyle (h)}^{\alpha} {}^{0}\nabla_{\alpha} 
\chi_2 \bigr) \Bigr] = 0 \; , 
\label{eqn:ExistCondExtSym} 
\end{equation} 
where $\chi$ is defined by Eq.(\ref{eqn:ChiDef}). 
Since a horizon two-sphere in the background spacetime 
is spherically symmetric, 
it will be reasonable to take into account all modes of perturbations of 
a horizon two-sphere on an equal footing and allow 
$A^{\scriptscriptstyle (i)}(\theta, \phi)$ to take arbitrary forms.  
Then, we can show from Eq.(\ref{eqn:ExistCondExtSym}) 
that $\chi$ should be written in the form   
\begin{equation} 
\chi = F(u) \: \xi_{\scriptscriptstyle (i)}^{\lambda} 
{}^{0}\nabla_{\lambda} A^{\scriptscriptstyle (i)}(\theta, \phi) 
+ S(\theta, \phi) \; ,  
\label{eqn:ChiConditionExt} 
\end{equation} 
where $F(u)$ is a universal function of $u$ in the sense 
that it takes the same form for all $\varsigma^{\mu}$, 
while $S(\theta, \phi)$ denotes an arbitrary function that can vary 
from one extended symmetry vector to another. 

We notice that the term proportional to $F(u)$ in 
Eq.(\ref{eqn:ChiConditionExt}), which results from the fact that  
$\omega(u, \theta, \phi)$ and $Z^{\mu}$ can depend on 
the null coordinate, does not contribute to the central term 
$\mbox{\large $K$}[\varsigma_{1}, \varsigma_{2}]$. 
The functions 
$A^{\scriptscriptstyle (i)}(\theta, \phi)$ and $S(\theta, \phi)$, 
which depend only on the angular coordinates, are essential for 
contribution to $\mbox{\large $K$}[\varsigma_{1}, \varsigma_{2}]$. 
On the other hand, in order to obtain the Virasoro algebra, 
whether or not it has non-trivial central charges, 
we need the $\mathit{Diff}(S^1)$ group, i.e., 
the diffeomorphism group of a circle. 
Those facts imply that reduction of the group of 
the extended symmetries, which arbitrarily depends on the angular 
coordinates of a two-sphere,  
is necessary to obtain the Virasoro algebra 
with non-trivial central charges. 
Especially, since $A^{\scriptscriptstyle (i)}(\theta, \phi)$ describe 
diffeomorphisms on a spherically symmetric two-sphere, we must 
reduce the diffeomorphism group of a two-sphere to that of a circle. 
Of course, there are infinitely many ways to specify a circle on a 
two-sphere, and we do not have any convincing reasons to pick up 
a particular circle on a spherically symmetric two-sphere. 
Therefore, we should unnaturally reduce the the group of 
the extended symmetries 
in order to obtain the Virasoro algebra with non-trivial central charges, 
and such reduction will not be justified 
when we respect the spherical symmetry of a horizon two-sphere 
in the background spacetime.

\section{Summary and discussions} 
\label{sec:Discussion} 
\eqnum{0} 

In this paper, we have analyzed the asymptotic symmetries on Killing 
horizons in static and spherically symmetric spacetimes. 
The asymptotic symmetries are generated by asymptotic Killing vectors, 
along which the Lie derivatives of the perturbed metrics vanish on 
the Killing horizon in the background spacetime. 
We gave the general form of asymptotic Killing vectors 
and found that the asymptotic symmetry group consists of 
rigid $O(3)$ rotations of a horizon two-sphere and 
supertranslations along the null direction, 
which arbitrarily depend not only on the angular coordinates, 
but also on the null coordinate. 
The supertranslations which depend on the null coordinate 
will be expected to play an important role in physics of the black hole 
entropy, because supertranslations of the B.M.S. group also depend on 
the angular coordinates, with which no entropy will be associated. 
In particular, we found that the Lie bracket algebra of 
the supertranslations that depend only on the null coordinate 
is isomorphic to the $\mathit{Diff}(S^1)$ algebra. 
The existence of the $\mathit{Diff}(S^1)$ subalgebra was necessary, 
if the Bekenstein--Hawking formula could be reproduced 
for generic black holes by the same method as 
that of the B.T.Z. black hole. 
We also considered reduction of the asymptotic symmetry group, 
and discussed the difference between the structure of 
the asymptotic symmetry group on the cosmological horizon 
in the de Sitter spacetime and that of 
the B.M.S. group, by comparing with the isometry groups in the maximally 
symmetric spacetimes. 

We then introduced the notion of asymptotic Killing horizons, 
which are defined in perturbed spacetimes 
and possess the local properties similar to those of 
exact Killing horizons. 
Although it is trivial that the local properties of 
asymptotic Killing horizons are invariant under diffeomorphisms, 
we showed that those are also preserved under the 
two types of non-trivial transformations, which are equivalent 
to each other up to diffeomorphisms. 
Under dynamical field  transformations, 
the metric is transformed by the Lie derivatives along 
asymptotic Killing vectors, while the horizon generating vector is fixed. 
On the other hand, horizon deformations transform 
the horizon generating vector, with the metric fixed. 
Thus, we found that the asymptotic symmetry group generates 
deformations of an asymptotic Killing horizon, 
which keep its local structures. 
We also found that the supertranslations 
that depend non-trivially on the null coordinate 
play the distinguishable role in the behavior of the surface 
gravity of an asymptotic Killing horizon, 
based on which we speculated that 
those may be related to the Euclidean approach to 
black hole thermodynamics. 

In order to clarify the features of asymptotic Killing horizons, 
here we compare asymptotic Killing horizons with 
weakly isolated horizons\cite{WeaklyIsolated,DreyerEtAl}, 
which have been generalized from isolated horizons so that 
distorted and rotating horizons are incorporated. 
A weakly isolated horizon is defined 
by a pair $( \Delta, [l^{\mu}] )$ of a 3-d null hypersurface $\Delta$ representing the horizon and an equivalence class $[l^{\mu}]$ 
of null vectors on $\Delta$ under constant rescalings. 
Those null vectors $l^{\mu} \in [l^{\mu}]$ generate the translational 
isometry along the null direction on a weakly isolated horizon, and thus 
play a similar role to the horizon generating vector 
$\xi_{\scriptscriptstyle (h)}^{\mu}$ of an asymptotic Killing horizon. 
On an asymptotic Killing horizon, we can show 
by using Eq.(\ref{eqn:deltaGamma}) that 
\begin{equation} 
\xi^{\mu}_{\scriptscriptstyle (h)} 
\bigl[ {\cal L}_{\zeta} , \nabla_{\mu} \bigr] 
\xi^{\nu}_{\scriptscriptstyle (h)} = 
\xi^{\mu}_{\scriptscriptstyle (h)} \, \xi^{\rho}_{\scriptscriptstyle (h)} 
\delta \Gamma^{\nu}_{\rho \mu} = \overline{\delta} \kappa \: 
\xi^{\nu}_{\scriptscriptstyle (h)} \; , 
\label{eqn:PartialSymmetry} 
\end{equation} 
which does not vanish even if $\zeta^{\mu}$ is equal to 
$\xi_{\scriptscriptstyle (h)}^{\mu}$, because 
$\xi_{\scriptscriptstyle (h)}^{\mu}$ is not an exact Killing vector 
in perturbed spacetimes, and then $\overline{\delta} \kappa$ does not 
vanish in general. 
On the contrary, one of the conditions of weakly isolated horizons 
requires 
\begin{equation} 
l^{\mu} \bigl[ {\cal L}_{l} , \nabla_{\mu} \bigr] l^{\nu} = 0 \; . 
\label{eqn:WeakIsoCond} 
\end{equation} 
In fact, the null vectors $l^{\mu}$ that 
satisfy Eq.(\ref{eqn:WeakIsoCond}) do not 
exist on an asymptotic Killing horizon generally. Therefore, 
asymptotic Killing horizons do not necessarily satisfy the condition of 
weakly isolated horizons, and the asymptotic symmetry group generates deformations of a Killing horizon that cannot be described by 
weakly isolated horizons. Correspondingly, the symmetry group of 
a weakly isolated horizon does not contain the supertranslations 
that depend on the null coordinate\cite{WeaklyIsolated}.  
This is related to the fact that the surface gravity is not constant 
over an asymptotic Killing horizon, while it is on a weakly 
isolated horizon (the zero-th law of weakly isolated horizons). 
However, as we mentioned before, 
the variation of the surface gravity over an asymptotic Killing horizon 
does not necessarily violate the zero-th law of 
ordinary black hole thermodynamics, because 
the surface gravity of an asymptotic Killing horizon is not 
directly related to the temperature of the horizon. 
Accordingly, we have not analyze the first law of 
asymptotic Killing horizons either, 
whereas the first law of weakly isolated horizons has been established. 
Another remarkable difference is that asymptotic Killing horizons 
are defined only when an exact Killing horizon exists 
in the background spacetime, while weakly isolated horizons 
can be defined without referring to any background structures. 

By employing the covariant phase space formalism and focusing on 
the Einstein theory with or without the cosmological constant, 
we also considered the Poisson bracket algebra of the conserved charges 
conjugate to asymptotic Killing vectors, and showed 
that the Poisson bracket algebra does not acquire non-trivial central 
charges, similarly to the result in the case of 
weakly isolated horizons\cite{DreyerEtAl}. 
Hence, the Poisson bracket algebra of 
the asymptotic symmetry group cannot contain 
the Virasoro algebra with non-trivial central charges, 
while the Lie bracket algebra contains the 
$\mathit{Diff}(S^1)$ algebra as a natural subalgebra. 
Therefore, we see that we cannot derive 
the Bekenstein--Hawking formula 
from the asymptotic symmetries on Killing horizons
by the same method as that of the B.T.Z. black hole. 
We then considered the weakened boundary condition and 
derived the extended symmetries, 
where null vectors on the Killing horizon in the background 
spacetime remain null in perturbed spacetimes. 
We found that the group of the extended symmetries that can result in 
a central charge in the Poisson bracket algebra 
is essentially described by the functions dependent 
only on the angular coordinates 
of a spherically symmetric two-sphere. 
Hence, unnatural reduction of that group 
to the $\mathit{Diff}(S^1)$ subgroup is necessary in order 
to obtain the Virasoro algebra with non-trivial central charges. 
It involves an artificial choice 
of a circle among infinitely many possible ones on a two-sphere, 
and thus it will not be justified because a horizon two-sphere 
in the background spacetime is spherically symmetric. 

Because the Virasoro algebra is associated with conformal invariance 
and event horizons are conformally invariant,  
it will be meaningful to describe the asymptotic symmetries 
on Killing horizons in terms of conformal transformations. 
(The B.M.S. group is also related to conformal completion of 
an asymptotically flat spacetime.) 
We first notice that the asymptotic symmetry group on a Killing horizon 
is given only by the subgroup of conformal transformations 
whose conformal factors are unity on the horizon. 
When we allow the conformal factors to take arbitrary forms on the 
horizon, it turns out, from the existence condition of 
well-defined conserved charges, that 
the conformal factors are required to have 
universal dependence on the null coordinate on the horizon, 
while they can depend arbitrarily on 
the angular coordinates. 
This is the same circumstance as that we are faced with 
in the analysis of the extended symmetries. 
We need to unnaturally reduce the group 
in order to obtain the Virasoro algebra, even if it is possible. 
Therefore, if we attempt to derive the Virasoro algebra, 
it is natural to consider only the conformal 
transformations whose conformal factors are unity on the horizon, 
whereas we cannot obtain non-trivial central charges. 

In this paper, we have presented the explicit results only for 
the future cosmological horizon in the 4-d de Sitter spacetime, 
but those results are valid for more general cases. 
The analyses in Section \ref{sec:AsymptoticSymmetries} and 
Section \ref{sec:AsymptoticKillingHorizon}, except for 
the reduced asymptotic symmetries, 
apply to arbitrary Killing horizons with the non-vanishing surface 
gravity in static and spherically symmetric 4-d spacetimes. 
If those spacetimes are described by the Einstein theory with or 
without the cosmological constant, the results in 
Section \ref{sec:PoissonBracket} also hold 
without additional restrictions. 
Furthermore, it will be straightforward to generalize 
those results into higher-dimensional spacetimes, 
as it is obvious from the derivations of 
asymptotic Killing vectors and 
the central term of the Poisson bracket algebra. 
The asymptotic symmetry group will consist of supertranslations 
in the null direction and rigid rotations of 
a horizon sphere with the higher-dimensional spherical symmetries. 
We will obtain the same conclusion about reduction of the group of 
the extended symmetries as well. 
It is also expected that supertranslations and 
the rotation around the symmetry axis are allowed as asymptotic 
symmetries even in the case of 4-d rotating black holes.
In this case, however, 
we can pick up the direction of the rotational symmetry as the 
special direction on a two-sphere, and then we may naturally reduce 
the group of the extended symmetries to 
the $\mathit{Diff}(S^1)$ subgroup. 
Moreover, we need not worry about reduction 
in the case of 3-d black holes, where a horizon sphere is 
a circle. In this respect, it will be interesting to clarify whether 
the extended symmetries are related to the microscopic 
derivations of the entropy in the 
3-d spacetimes\cite{CarlipAdS,MaldacenaStrominger}. 
However, it does not provide the general framework, 
since such a specific choice of a direction does not work 
in spherically symmetric spacetimes 
with the dimensions greater than four. 
It does not work either in the case of 
higher-dimensional rotating black holes, because we will have two or more directions of rotational symmetry. 
Therefore, we can conclude that the asymptotic symmetries on Killing 
horizons or their extension do not provide the universal method, 
by which the Bekenstein--Hawking formula is reproduced 
for generic black holes in the same way as that of the B.T.Z. black hole. 

However, it does not necessarily imply that we cannot describe the 
black hole entropy by the asymptotic symmetries on Killing horizons. 
In fact, there are some evidences that asymptotic symmetries 
are important to understand the black hole entropy, such as 
the successful results in the 
3-d spacetimes\cite{Strominger,CarlipAdS,MaldacenaStrominger} 
and the recent investigation of the asymptotic symmetries 
on the spacelike hypersurfaces at the infinite past and future 
in the de Sitter spacetime\cite{Strominger2}. 
Since the universal feature 
in black hole spacetimes is presence of black hole horizons, 
we can expect that 
the asymptotic symmetries on Killing horizons are responsible for 
the black hole entropy. 
In addition, 2-d conformal symmetry plays a crucial role 
in the successful results mentioned above and 
the analysis of Ref.\cite{Solodukhin}, 
as well as in string theory. 
Presence of the $\mathit{Diff}(S^1)$ subgroup, 
which results from the supertranslations dependent 
only on the null coordinate, indicates that 
the asymptotic symmetries on Killing horizons 
are also equipped with the feature related to 2-d conformal symmetry. 
Conformal symmetry on black hole horizons has been analyzed also  
in Ref.\cite{SachsSolo} by the method based on the optical metric, 
while the relevant asymptotic symmetries in Ref.\cite{SachsSolo} are 
essentially described by constant parameters and 
a function dependent only on the angular coordinates, 
and hence those are different from 
the asymptotic symmetries considered in this paper. 

Although the Poisson bracket algebra of the asymptotic symmetry group 
on Killing horizons does not possess a central charge, 
we may possibly reproduce the Bekenstein--Hawking formula by 
different methods from that of the B.T.Z. black hole. 
Particularly, we will need to consider quantum theories, 
while the analyses in this paper are purely classical.   
The commutators of the generators in a quantum theory 
may acquire a central charge, 
as in the case of quantization of a string, whereas  
such a quantum theory has not been developed so far.  
It will be also interesting to explore the possibility that 
we derive the temperature of a black hole horizon 
from the asymptotic symmetries. 
In any case, we should clarify in future investigations 
whether we can indeed derive 
the thermodynamic features of black holes microscopically, 
by methods based on the asymptotic symmetries on Killing horizons.

\vspace{0.5cm}
{{\bf Acknowledgment}}

I would like to thank S. Carlip for helpful discussions. 
I also thank G. W. Gibbons and T. Maki for discussions 
at the early stage of this work, 
and K. Maeda for continuous encouragement. This work is supported 
by the Grant-in-Aid for Encouragement of Young Scientists 
from the Japanese Ministry of Education, Science, Sports 
and Culture (No.13740162), and from Waseda University. 


\appendix
\section{Derivation of asymptotic Killing vectors} 
\label{sec:AsymKillEq} 
\eqnum{0} 

In this appendix, we write down the explicit forms of 
Eqs.(\ref{eqn:AsymKillingAbs}) and (\ref{eqn:AsymKillingSecond}), 
and derive asymptotic Killing vectors $\zeta^{\mu}$ and 
reduced asymptotic Killing vectors $\eta^{\mu}$. 
To do so, we write the asymptotic forms of the metric components as 
\begin{eqnarray}
g_{u v} \! & \! \! \! = \! \! \! & \! 
- 2 + v \; h_{u v}(u, \theta, \phi) 
+ v^2 \: k_{u v}(u, \theta, \phi) + {\cal O}(v^{3}) \; , 
\label{eqn:AsyMetricuv} \\
g_{\theta \theta} \! & \! \! \! = \! \! \! & \! 
r_{\scriptscriptstyle H}^2  
+ v \; h_{\theta \theta}(u, \theta, \phi) 
+ v^2 \: k_{\theta \theta}(u, \theta, \phi) + {\cal O}(v^{3}) \; , 
\label{eqn:AsyMetrictt} \\
g_{\phi \phi} \! & \! \! \! = \! \! \! & \! 
r_{\scriptscriptstyle H}^2 \sin^2 \! \theta 
+ v \; h_{\phi \phi}(u, \theta, \phi) 
+ v^2 \: k_{\phi \phi}(u, \theta, \phi) + {\cal O}(v^{3}) \; ,  
\label{eqn:AsyMetricpp} \\ 
g_{\mu \nu} \! & \! \! \! = \! \! \! & \! 0 
+ v \; h_{\mu \nu}(u, \theta, \phi) 
+ v^2 \: k_{\mu \nu}(u, \theta, \phi) + {\cal O}(v^{3}) 
~~~~ \mbox{ for other components} \; ,  
\label{eqn:AsyMetricelse} 
\end{eqnarray} 
where the leading terms are specified from 
Eq.(\ref{eqn:BackMetricAsym}). 

We first leave $h_{\mu \nu}$ arbitrary and derive asymptotic Killing vector $\zeta^{\mu}$. By substituting 
Eqs.(\ref{eqn:AsyMetricuv})--(\ref{eqn:AsyMetricelse}) and 
(\ref{eqn:ZetaExpand}) into Eq.(\ref{eqn:AsymKillingAbs}), 
and requiring that the terms of ${\cal O}(1)$ in 
Eq.(\ref{eqn:AsymKillingAbs}) vanish, we have 
\begin{eqnarray}
& & h_{v v} \; \zeta^{v}_{\scriptscriptstyle (0)} - 4 \:  
\zeta^{u}_{\scriptscriptstyle (1)} = 0 \; , 
\label{eqn:AsyKill1DSvv}  \\ 
& & h_{u v} \; \zeta^{v}_{\scriptscriptstyle (0)} - 2 \: 
\zeta^{v}_{\scriptscriptstyle (1)} - 2 \: \partial_{u} 
\zeta^{u}_{\scriptscriptstyle (0)} = 0 \; , 
\label{eqn:AsyKill1DSuv}  \\ 
& & h_{u u} \; \zeta^{v}_{\scriptscriptstyle (0)} - 4 \: 
\partial_{u} \zeta^{v}_{\scriptscriptstyle (0)} = 0 \; , 
\label{eqn:AsyKill1DSuu}  \\ 
& & h_{v \theta} \; \zeta^{v}_{\scriptscriptstyle (0)} - 2 \: 
\partial_{\theta} \zeta^{u}_{\scriptscriptstyle (0)} + 
r_{\scriptscriptstyle H}^2 \, \zeta^{\theta}_{\scriptscriptstyle (1)}  
= 0 \; , 
\label{eqn:AsyKill1DSvt}  \\ 
& & h_{v \phi} \; \zeta^{v}_{\scriptscriptstyle (0)} - 2 \: 
\partial_{\phi} \zeta^{u}_{\scriptscriptstyle (0)} + 
r_{\scriptscriptstyle H}^2 \sin^2 \! \theta \; 
\zeta^{\phi}_{\scriptscriptstyle (1)} = 0 \; , 
\label{eqn:AsyKill1DSvp}  \\ 
& & h_{u \theta} \; \zeta^{v}_{\scriptscriptstyle (0)} - 2 \: 
\partial_{\theta} \zeta^{v}_{\scriptscriptstyle (0)} 
+ r_{\scriptscriptstyle H}^2 \, 
\partial_{u} \zeta^{\theta}_{\scriptscriptstyle (0)} = 0 \; , 
\label{eqn:AsyKill1DSut}  \\
& & h_{u \phi} \; \zeta^{v}_{\scriptscriptstyle (0)} - 2 \: 
\partial_{\phi} \zeta^{v}_{\scriptscriptstyle (0)} 
+ r_{\scriptscriptstyle H}^2 
\sin^2 \! \theta \; \partial_{u} \zeta^{\phi}_{\scriptscriptstyle (0)} 
= 0 \; , 
\label{eqn:AsyKill1DSup}  \\ 
& & h_{\theta \theta} \; \zeta^{v}_{\scriptscriptstyle (0)} 
+ 2 \: r_{\scriptscriptstyle H}^2 \, 
\partial_{\theta} \zeta^{\theta}_{\scriptscriptstyle (0)} = 0 \; , 
\label{eqn:AsyKill1DStt}  \\ 
& & h_{\theta \phi} \; \zeta^{v}_{\scriptscriptstyle (0)} 
+ r_{\scriptscriptstyle H}^2 \, 
\partial_{\phi} \zeta^{\theta}_{\scriptscriptstyle (0)} 
+ r_{\scriptscriptstyle H}^2 
\sin^2 \! \theta \; \partial_{\theta} \zeta^{\phi}_{\scriptscriptstyle (0)} 
= 0 \; , 
\label{eqn:AsyKill1DStp}  \\ 
& & h_{\phi \phi} \; \zeta^{v}_{\scriptscriptstyle (0)} 
+ 2 \: r_{\scriptscriptstyle H}^2 
\sin \theta \cos \theta \; \zeta^{\theta}_{\scriptscriptstyle (0)} 
+ 2 \: r_{\scriptscriptstyle H}^2 \sin^2 \! \theta \; \partial_{\phi} 
\zeta^{\phi}_{\scriptscriptstyle (0)}  = 0 \; . 
\label{eqn:AsyKill1DSpp}  
\end{eqnarray} 
Since Eqs.(\ref{eqn:AsyKill1DSvv})--(\ref{eqn:AsyKill1DSpp}) should 
hold for arbitrary forms of $h_{\mu \nu}$, we obtain 
\begin{eqnarray} 
& & \zeta^{v}_{\scriptscriptstyle (0)} = 0 \; ,  
\label{eqn:ZetaCondvv} \\  
& & \zeta^{u}_{\scriptscriptstyle (1)} = 0 \; , 
\label{eqn:ZetaConduv} \\ 
& & \zeta^{v}_{\scriptscriptstyle (1)} = - \; \partial_{u} 
\zeta^{u}_{\scriptscriptstyle (0)} \; , 
\label{eqn:ZetaConduu} \\ 
& & \zeta^{\theta}_{\scriptscriptstyle (1)} = 
\frac{2}{r_{\scriptscriptstyle H}^2} 
\; \partial_{\theta} \zeta^{u}_{\scriptscriptstyle (0)} \; , 
\label{eqn:ZetaCondvt} \\ 
& & \zeta^{\phi}_{\scriptscriptstyle (1)} = 
\frac{2}{r_{\scriptscriptstyle H}^2 \sin^2 \! \theta} \; \partial_{\phi} 
\zeta^{u}_{\scriptscriptstyle (0)} \; , 
\label{eqn:ZetaCondvp} \\ 
& & \partial_{u} \zeta^{\theta}_{\scriptscriptstyle (0)} = 0 \; , 
\label{eqn:ZetaCondut} \\ 
& & \partial_{u} \zeta^{\phi}_{\scriptscriptstyle (0)} = 0 \; , 
\label{eqn:ZetaCondup} \\ 
& & \partial_{\theta} \zeta^{\theta}_{\scriptscriptstyle (0)} = 0 \; ,  
\label{eqn:ZetaCondtt} \\ 
& & \partial_{\phi} \zeta^{\theta}_{\scriptscriptstyle (0)}  
+ \sin^2 \! \theta \; 
\partial_{\theta} \zeta^{\phi}_{\scriptscriptstyle (0)} = 0 \; , 
\label{eqn:ZetaCondtp} \\ 
& & \cot \theta \; \zeta^{\theta}_{\scriptscriptstyle (0)} + 
\partial_{\phi} \zeta^{\phi}_{\scriptscriptstyle (0)} = 0 \; . 
\label{eqn:ZetaCondpp} 
\end{eqnarray} 
We notice that Eqs.(\ref{eqn:ZetaCondut})--(\ref{eqn:ZetaCondpp}) 
decouple from the others and give 
the 2-d Killing equation on a sphere for 
$\zeta^{\theta}_{\scriptscriptstyle (0)}$ and 
$\zeta^{\phi}_{\scriptscriptstyle (0)}$, while 
Eqs.(\ref{eqn:ZetaCondvv})--(\ref{eqn:ZetaCondvp}) 
determine the other variables in terms of 
$\zeta^{u}_{\scriptscriptstyle (0)}$. 
When we solve those equations, 
we find that the general form of asymptotic Killing vectors 
$\zeta^{\mu}$ is written as a linear combination of 
\begin{eqnarray} 
\zeta^{\mu}_{\scriptscriptstyle (\mbox{\scriptsize s})}
 \! & \! \! \! = \! \! \! & \! \Bigl[ - v \; \partial_{u} 
\zeta^{u}_{\scriptscriptstyle (0)}+ {\cal O}(v^2) \Bigr] 
\frac{\partial}{\partial v} + \Bigl[ \zeta^{u}_{\scriptscriptstyle (0)} 
+ {\cal O}(v^2) \Bigr] 
\frac{\partial}{\partial u} + \Bigl[ 
\frac{2}{r_{\scriptscriptstyle H}^2} \; v \; \partial_{\theta} 
\zeta^{u}_{\scriptscriptstyle (0)} + {\cal O}(v^2) \Bigr] 
\frac{\partial}{\partial \theta} 
\nonumber \\ & & 
+ \; \Bigl[ 
\frac{2}{r_{\scriptscriptstyle H}^2 \sin^2 \! \theta} \; v \; 
\partial_{\phi} 
\zeta^{u}_{\scriptscriptstyle (0)} + {\cal O}(v^2) 
\Bigr] \frac{\partial}{\partial \phi} \; ,  
\end{eqnarray}  
and the Killing vectors $\xi^{\mu}_{\scriptscriptstyle (i)}$ 
of $O(3)$ rotations. 
Then, $\zeta^{\mu}$ is rewritten, 
by using Eqs.(\ref{eqn:omegafuncDef}) and (\ref{eqn:XDef}), as 
\begin{equation}
\zeta^{\mu} = \omega(u,\theta,\phi) \; \xi^{\mu}_{\scriptscriptstyle (h)} 
+ a^{\scriptscriptstyle (i)} \, \xi^{\mu}_{\scriptscriptstyle (i)} 
+ v \, X^{\mu}(\omega) + v^2 \, Y^{\mu} + {\cal O}(v^3) \; .  
\end{equation}

Next, we derive reduced asymptotic Killing vectors $\eta^{\mu}$ 
in the de Sitter spacetime. We substitute 
Eqs.(\ref{eqn:AsyMetricuv})--(\ref{eqn:AsyMetricelse}) as well as 
Eq.(\ref{eqn:EtaExpand}) into Eq.(\ref{eqn:AsymKillingSecond}), 
with $r_{\scriptscriptstyle H} = \ell$ and $h_{\mu \nu}$ set equal to 
that of the de Sitter spacetime, as 
\begin{equation} 
h_{u v} =  - \, \frac{4 \, u}{\ell^2} \; , 
~ 
h_{\theta \theta} =  4 \, u \; ,  
~ 
h_{\phi \phi} =  4 \, u \, \sin^2 \theta \; , 
~   
h_{\mu \nu} = 0 \; ~ \mbox{for other components} \; .  
\end{equation} 
Now,  Eq.(\ref{eqn:AsymKillingSecond}) is required to hold for arbitrary 
forms of $k_{\mu \nu}$. 
From the condition that the terms of ${\cal O}(1)$ in 
Eq.(\ref{eqn:AsymKillingSecond}) vanish, 
we obtain the same equations as 
Eqs.(\ref{eqn:ZetaCondvv}) --(\ref{eqn:ZetaCondpp}) with $\zeta^{\mu}$ replaced by $\eta^{\mu}$. By making use of those equations, 
the condition that the terms of ${\cal O}(v)$ vanish yields 
\begin{eqnarray} 
& & \eta^{u}_{\scriptscriptstyle (2)} = 0 \; , 
\label{eqn:EtaCondvv} \\ 
& & \eta^{v}_{\scriptscriptstyle (2)} = \frac{u}{\ell^2} \: \partial_{u} 
\eta^{u}_{\scriptscriptstyle (0)} - \frac{1}{\ell^2} \: 
\eta^{u}_{\scriptscriptstyle (0)} \; , 
\label{eqn:EtaConduv} \\ 
& & \eta^{\theta}_{\scriptscriptstyle (2)} = - \, \frac{2 \: u}{\ell^4} \: 
\partial_{\theta} \eta^{u}_{\scriptscriptstyle (0)} \; , 
\label{eqn:EtaCondvt} \\  
& & \eta^{\phi}_{\scriptscriptstyle (2)} = - \, 
\frac{2 \: u}{\ell^4 \sin^2 \theta} \: 
\partial_{\phi} \eta^{u}_{\scriptscriptstyle (0)} \; , 
\label{eqn:EtaCondvp} \\ 
& & \partial_{\theta} \, \partial_{u} \eta^{u}_{\scriptscriptstyle (0)} 
= 0 \; , 
\label{eqn:EtaCondut} \\ 
& & \partial_{\phi} \, \partial_{u} \eta^{u}_{\scriptscriptstyle (0)} 
= 0 \; , 
\label{eqn:EtaCondup} \\ 
& & \eta^{u}_{\scriptscriptstyle (0)}  - u \: \partial_{u} 
\eta^{u}_{\scriptscriptstyle (0)} + \partial_{\theta} \, \partial_{\theta} 
\eta^{u}_{\scriptscriptstyle (0)} = 0 \; , 
\label{eqn:EtaCondtt} \\ 
& & \partial_{\theta} \, \partial_{\phi} \eta^{u}_{\scriptscriptstyle (0)} 
= \cot \theta \; \partial_{\phi} \eta^{u}_{\scriptscriptstyle (0)} \; , 
\label{eqn:EtaCondtp} \\ 
& & \partial_{\theta} \, \partial_{\theta} 
\eta^{u}_{\scriptscriptstyle (0)} 
- \cot \theta \; \partial_{\theta} \eta^{u}_{\scriptscriptstyle (0)} 
- \frac{1}{\sin^2 \theta} \; \partial_{\phi} \, \partial_{\phi} 
\eta^{u}_{\scriptscriptstyle (0)} = 0 \; ,   
\label{eqn:EtaCondpp} 
\end{eqnarray}  
and $\partial_{u} \eta^{v}_{\scriptscriptstyle (1)} = 0$. 
Eqs.(\ref{eqn:EtaCondut})--(\ref{eqn:EtaCondpp}) constrain 
$\eta^{u}_{\scriptscriptstyle (0)}$ to take the forms 
\begin{equation} 
\eta^{u}_{\scriptscriptstyle (0)} = 
u \; , ~~ \sin \theta \; e^{- i \phi} \; , ~~ \cos \theta \; ,  
~~ \sin \theta \; e^{i \phi} \;  ~~ \mbox{or} ~~ 0,  
\end{equation} 
while Eqs.(\ref{eqn:EtaCondvv})--(\ref{eqn:EtaCondvp}) determine 
 $\eta^{u}_{\scriptscriptstyle (2)}$, 
$\eta^{v}_{\scriptscriptstyle (2)}$, 
$\eta^{\theta}_{\scriptscriptstyle (2)}$ and 
$\eta^{\phi}_{\scriptscriptstyle (2)}$ in terms of 
$\eta^{u}_{\scriptscriptstyle (0)}$. 
Then, by using again the same 
equations as Eqs.(\ref{eqn:ZetaCondvv})--(\ref{eqn:ZetaCondpp}) 
with $r_{\scriptscriptstyle H} = \ell$, 
we can derive the general form of $\eta^{\mu}$, which is given 
as a linear combination of the vectors defined by  
Eqs.(\ref{eqn:AsymKillVecSec01})--(\ref{eqn:AsymKillVecSec34}). 

\section{Calculation of a central charge} 
\label{sec:CalcCenter} 
\eqnum{0} 

In order to find the form of the central term 
$\mbox{\large $K$}[\varsigma_{1}, \varsigma_{2}]$ in 
Eq.(\ref{eqn:CenterExtd}), we need to calculate 
\begin{equation} 
\overline{\delta}_{\varsigma_{2}} 
\mbox{\large ${\cal H}$}[g; \varsigma_{1}] 
= \int_{\partial {\cal C}} \frac{1}{16 \pi} \, 
\varepsilon_{\beta \alpha \mu \nu} \: c^{\beta \alpha} \; , 
\label{eqn:CentralChargeEvalAp} 
\end{equation} 
where $c^{\beta \alpha}$ is defined as
\begin{equation} 
c^{\beta \alpha} = 
\bigl( \nabla^{\gamma} \varsigma^{\alpha}_{1} \bigr) 
 \bigl( \nabla_{\gamma} \varsigma^{\beta}_{2} \bigr) 
- \frac{1}{2} \, R^{\beta \alpha}_{~~~ \gamma \lambda} \, 
\varsigma^{\gamma}_{1} \varsigma^{\lambda}_{2} 
+ \frac{1}{2} \: \varsigma^{[\beta}_{1} \varsigma^{\alpha]}_{2} 
\bigl( R - 2 \Lambda \bigr) 
+ \bigl( \nabla^{[ \alpha} \varsigma^{\beta ]}_{1} \bigr) 
\bigl( \nabla_{\gamma} \varsigma^{\gamma}_{2} \bigr) 
- \bigl( 1 \leftrightarrow 2 \bigr) \; ,  
\label{eqn:CentralIntgrand} 
\end{equation} 
for extended symmetry vectors $\varsigma^{\mu}$. 
Eq.(\ref{eqn:CentralChargeEvalAp}) along with 
Eq.(\ref{eqn:CentralIntgrand}) reduces to Eq.(\ref{eqn:CentralChargeEin}) 
when asymptotic Killing vectors $\zeta^{\mu}$ are substituted in place 
of $\varsigma^{\mu}$. 

As we argued in Section \ref{sec:PoissonBracket}, we only have to 
evaluate Eq.(\ref{eqn:CentralChargeEvalAp}) on the background 
spacetime. Then, we can apply the exact Killing equations for 
$\xi_{\scriptscriptstyle (h)}^{\mu}$ and 
$\xi^{\mu}_{\scriptscriptstyle (i)}$, which are valid globally in the 
background spacetime. 
We also introduce $\Delta_{\mu \nu} \equiv 
{\cal L}_{\varsigma} g_{\mu \nu} |_{v = 0}$, 
which is expressed, in terms of the Killing vectors, as 
\begin{equation} 
\Delta^{\mu \nu} = 2 \, \xi_{\scriptscriptstyle (h)}^{( \nu} 
\bigl( \nabla^{\mu )} \omega \bigr) 
+ 2 \, \xi^{( \nu}_{\scriptscriptstyle (i)} 
\bigl( \nabla^{\mu )} A^{\scriptscriptstyle (i)} \bigr) 
+ 2 \, Z^{( \nu} \bigl( \nabla^{\mu )} v \bigr) + {\cal O}(v) \; . 
\label{eqn:DeltaDef} 
\end{equation} 
Here and hereafter, we omit the symbol $0$ 
that has been used for quantities in the background spacetime, 
because of simplicity of expressions. 

By substituting Eqs.(\ref{eqn:ExtdAsymVec}) and (\ref{eqn:DeltaDef}) 
as well as the Killing equations into Eq.(\ref{eqn:CentralIntgrand}), 
we obtain, after a lengthy calculation 
involving integration by parts, 
\begin{eqnarray} 
c^{\beta \alpha} 
\! & \! \! \! = \! \! \! & \! 
\Bigl[ 2 \, \bigl( \nabla_{\gamma} \omega_1 \bigr) \bigl( 
\xi_{\scriptscriptstyle (h)}^{[ \alpha} \Delta^{\beta ] \gamma}_2 \bigr) 
+ \frac{\Delta_2}{2} \bigl\{ 
\xi_{\scriptscriptstyle (h)}^{[ \beta} 
\bigl( \nabla^{\alpha ]} \omega_1 \bigr) 
+ \xi^{[ \beta}_{\scriptscriptstyle (i)} 
\bigl( \nabla^{\alpha ]} A^{\scriptscriptstyle (i)}_1 \bigr) 
+ Z^{[ \beta}_1 \bigl( \nabla^{\alpha ]} v \bigr) \bigr\} 
\nonumber \\ 
\! & \! \! \! \! \! \! & \! 
+ \; 3 \, \bigl( \omega_1 \nabla_{\gamma} A^{\scriptscriptstyle (i)}_2 
\bigr) 
\bigl( \xi_{\scriptscriptstyle (h)}^{[ \beta} \nabla^{\gamma} 
\xi_{\scriptscriptstyle (i)}^{\alpha ]} 
+ \xi_{\scriptscriptstyle (i)}^{[ \beta} \nabla^{\gamma} 
\xi_{\scriptscriptstyle (h)}^{\alpha ]} \bigr) 
+ \frac{3}{2} \, \bigl(  A^{\scriptscriptstyle (i)}_1 \nabla_{\gamma} 
A^{\scriptscriptstyle (j)}_2 \bigr) 
\bigl( \xi_{\scriptscriptstyle (i)}^{[ \beta} 
\nabla^{\gamma} \xi_{\scriptscriptstyle (j)}^{\alpha ]} 
+ \xi_{\scriptscriptstyle (j)}^{[ \beta} 
\nabla^{\gamma} \xi_{\scriptscriptstyle (i)}^{\alpha ]} \bigr) 
\nonumber \\ 
\! & \! \! \! \! \! \! & \! 
+ \; 2 \, \bigl( \xi_{\scriptscriptstyle (h)}^{\gamma} \nabla_{\gamma} 
\omega_1 \bigr) \bigl( \xi_{\scriptscriptstyle (h)}^{[ \beta} 
\nabla^{\alpha ]} \omega_2 \bigr) 
+ 2 \, \bigl( \xi_{\scriptscriptstyle (i)}^{\gamma} \nabla_{\gamma} 
\omega_1 \bigr) \bigl( \xi_{\scriptscriptstyle (h)}^{[ \beta} 
\nabla^{\alpha ]} A^{\scriptscriptstyle (i)}_2 \bigr) 
+ 2 \, \bigl( \nabla_{\gamma} A^{\scriptscriptstyle (i)}_1 \bigr) 
\bigl( \nabla^{\gamma} v \bigr) \, 
\xi_{\scriptscriptstyle (i)}^{[ \alpha} Z^{\beta ]}_2 
\nonumber \\ 
\! & \! \! \! \! \! \! & \! 
+ \; \bigl( \nabla^{\gamma} A^{\scriptscriptstyle (i)}_1 \bigr) 
\bigl( \nabla_{\gamma} A^{\scriptscriptstyle (j)}_2 \bigr) \, 
\xi_{\scriptscriptstyle (i)}^{[ \alpha} 
\xi_{\scriptscriptstyle (j)}^{\beta ]} 
+ \bigl( \nabla_{\gamma} v \bigr) 
\bigl( \nabla^{\gamma} v \bigr) \, Z^{[ \alpha}_{1} 
Z^{\beta ]}_{2} - \bigl( 1 \leftrightarrow 2 \bigr) \Bigr] 
+ \nabla_{\gamma} F^{[ \beta \alpha \gamma ]} 
+ {\cal O}(v) \; , 
\label{eqn:ExpressionCAp} 
\end{eqnarray} 
where $\Delta \equiv g^{\mu \nu} \Delta_{\mu \nu}$ and 
$F^{[ \beta \alpha \gamma ]}$ is given by 
\begin{equation} 
F^{[ \beta \alpha \gamma ]} \equiv 
A^{\scriptscriptstyle (i)}_1 \omega_2 \: 
\xi_{\scriptscriptstyle (h)}^{[ \beta} \nabla^{\gamma} 
\xi_{\scriptscriptstyle (i)}^{\alpha ]} + \frac{1}{2} \, 
A^{\scriptscriptstyle (i)}_1 A^{\scriptscriptstyle (j)}_2 \, 
\xi_{\scriptscriptstyle (i)}^{[ \beta} 
\nabla^{\alpha} \xi_{\scriptscriptstyle (j)}^{\gamma ]} 
+ v \: \omega_1 \, Z^{[ \beta}_2 
\nabla^{\gamma} \xi_{\scriptscriptstyle (h)}^{\alpha ]} 
+ v \: A^{\scriptscriptstyle (i)}_1 \, 
Z^{[ \beta}_2 \nabla^{\gamma} 
\xi_{\scriptscriptstyle (i)}^{\alpha ]} 
- \bigl( 1 \leftrightarrow 2 \bigr) 
\; .  
\end{equation} 
In deriving Eq.(\ref{eqn:ExpressionCAp}), we assumed that 
$\varsigma^{\mu}$ are regular at $v = 0$, 
substituted the Einstein equation as well as 
the hypersurface orthogonality of 
$\xi_{\scriptscriptstyle (h)}^{\mu}$, and used the fact that 
$\nabla^{\mu} v$ is proportional to $\xi_{\scriptscriptstyle (h)}^{\mu}$ 
at $v = 0$. 

We then rewrite $\varepsilon_{\beta \alpha \mu \nu} \: c^{\beta \alpha}$ 
by using the relations 
$\xi_{\scriptscriptstyle (h)}^{\alpha} \nabla_{\alpha} 
A^{\scriptscriptstyle (i)} = 0$ and 
$\xi_{\scriptscriptstyle (h)}^{\gamma} \nabla_{\gamma} 
\xi_{\scriptscriptstyle (i)}^{\lambda} = 
\xi_{\scriptscriptstyle (i)}^{\gamma} \nabla_{\gamma} 
\xi_{\scriptscriptstyle (h)}^{\lambda} = 0$.  
The former is the condition for $\varsigma^{\mu}$, as we derived in 
Section \ref{sec:PoissonBracket}, and the latter is shown from 
the Lie brackets between the Killing vectors and the fact that 
a horizon two-sphere is spherically symmetric. 
By using again Eq.(\ref{eqn:DeltaDef}) and the hypersurface 
orthogonality of $\xi_{\scriptscriptstyle (h)}^{\mu}$, we have 
\begin{eqnarray} 
\varepsilon_{\beta \alpha \mu \nu} \: c^{\beta \alpha} 
\! & \! \! \! = \! \! \! & \! 
\varepsilon_{\mu \nu} \Bigl[ 2 \, \xi_{\scriptscriptstyle (h)}^{\lambda} 
\, \Delta_{2 \gamma \lambda}  \bigl( \nabla^{\gamma} \omega_1 \bigr) 
- \Delta_2 \; \xi_{\scriptscriptstyle (h)}^{\alpha} 
\nabla_{\alpha} \omega_1 
- \frac{\Delta_2}{2} \, \xi_{\scriptscriptstyle (i)}^{\alpha} 
\nabla_{\alpha} A^{\scriptscriptstyle (i)}_1 
- 2 \, \kappa \: \omega_1 \, \xi_{\scriptscriptstyle (i)}^{\gamma} 
\nabla_{\gamma} A^{\scriptscriptstyle (i)}_2 
- \bigl( 1 \leftrightarrow 2 \bigr) \Bigr] 
\nonumber \\ 
\! & \! \! \! \! \! \! & \! 
+ \frac{1}{3} \, d_{\mu} \, \varepsilon_{\beta \alpha \gamma \nu} 
F^{[ \beta \alpha \gamma ]} 
+ {\cal O}(v) 
\nonumber \\ 
\! & \! \! \! = \! \! \! & \! 
\varepsilon_{\mu \nu} \Bigl[ 
\bigl( \xi_{\scriptscriptstyle (i)}^{\alpha} 
\nabla_{\alpha} A^{\scriptscriptstyle (i)}_1 \bigr) 
\bigl( \xi_{\scriptscriptstyle (h)}^{\lambda} \nabla_{\lambda} \omega_2 
+ 2 \, \kappa \: \omega_2 - Z^{\lambda}_2 \nabla_{\lambda} v \bigr) 
- \bigl( \xi_{\scriptscriptstyle (i)}^{\alpha} 
\nabla_{\alpha} A^{\scriptscriptstyle (i)}_2  \bigr) 
\bigl( \xi_{\scriptscriptstyle (h)}^{\lambda} \nabla_{\lambda} \omega_1 
+ 2 \, \kappa \: \omega_1 - Z^{\lambda}_1 \nabla_{\lambda} v \bigr) 
\Bigr] 
\nonumber \\ 
\! & \! \! \! \! \! \! & \! 
+ \frac{1}{3} \, d_{\mu} \, \varepsilon_{\beta \alpha \gamma \nu} 
F^{[ \beta \alpha \gamma ]} 
+ {\cal O}(v) \; , 
\label{eqn:CentralEvalForm} 
\end{eqnarray} 
where $\varepsilon_{\mu \nu} = \varepsilon_{\beta \alpha \mu \nu} \; 
\xi_{\scriptscriptstyle (h)}^{\beta} \, n^{\alpha}$ 
is the volume element of a horizon two-sphere $\partial {\cal C}$, 
and the null vectors normal to $\partial {\cal C}$ are normalized as 
$\xi_{\scriptscriptstyle (h)}^{\beta} \, n_{\beta} = -1$. 
The second term of Eq.(\ref{eqn:CentralEvalForm}) 
does not make any contribution, because it is 
totally divergent and a horizon two-sphere has no boundary. 
Therefore, we obtain, 
\begin{equation} 
\overline{\delta}_{\varsigma_{2}} 
\mbox{\large ${\cal H}$}[g; \varsigma_{1}] = 
\int_{\partial {\cal C}} \frac{1}{16 \pi} \, 
\varepsilon_{\mu \nu} \Bigl[ 
\bigl( \xi_{\scriptscriptstyle (i)}^{\alpha} 
\nabla_{\alpha} A^{\scriptscriptstyle (i)}_1 \bigr) \, \chi_2 
- \bigl( \xi_{\scriptscriptstyle (i)}^{\alpha} 
\nabla_{\alpha} A^{\scriptscriptstyle (i)}_2  \bigr) \, \chi_1 \Bigr] 
\; , 
\label{eqn:CenterFin} 
\end{equation} 
where we defined as 
\begin{equation} 
\chi \equiv 
\xi_{\scriptscriptstyle (h)}^{\lambda} \nabla_{\lambda} \omega 
+ 2 \, \kappa \: \omega - Z^{\lambda} \, \nabla_{\lambda} v \; . 
\end{equation} 

We see that Eq.(\ref{eqn:CenterFin}) 
gives Eq.(\ref{eqn:CenterExtd}), and reduces to 
Eq.(\ref{eqn:VanishingC}) when $\varsigma^{\mu}$ are replaced by 
asymptotic Killing vectors $\zeta^{\mu}$. 

\section{Killing vectors in the de Sitter spacetime} 
\label{sec:KillingList} 
\eqnum{0} 

Here, we list the Killing vectors in the de Sitter spacetime, 
for the purpose of comparison with asymptotic Killing vectors. 
In the coordinate system of Eq.(\ref{eqn:DSKrsMetric}), 
those are given by 
\begin{eqnarray}
\xi^{\mu}_{\scriptscriptstyle (t)} 
\! & \! \! \! = \! \! \! & \! 
- \; v \; \frac{\partial}{\partial v} + 
u \; \frac{\partial}{\partial u} \; , 
\label{eqn:KillingDS01} \\ 
\xi^{\mu}_{\scriptscriptstyle (\bot)} 
\! & \! \! \! = \! \! \! & \! 
\ell \, \cos \theta \: \frac{\partial}{\partial v} - 
\frac{u^2}{\ell} \cos \theta \: \frac{\partial}{\partial u} 
- \frac{2 \: \ell \: u}{u v + \ell^2} \, \sin \theta \: 
\frac{\partial}{\partial \theta} \; , 
\label{eqn:KillingDSp2} \\ 
\xi^{\mu}_{\scriptscriptstyle (\parallel)} 
\! & \! \! \! = \! \! \! & \! 
\frac{v^2}{\ell} \cos \theta \: \frac{\partial}{\partial v} 
- \ell \, \cos \theta \: \frac{\partial}{\partial u} 
+ \frac{2 \: \ell \: v}{u v + \ell^2} \, 
\sin \theta \: \frac{\partial}{\partial \theta} \; , 
\label{eqn:KillingDSm2} \\ 
\xi^{\mu}_{\scriptscriptstyle (\bot\pm)} 
\! & \! \! \! = \! \! \! & \! 
\ell \, \sin \theta \; e^{\pm i \phi} \: 
\frac{\partial}{\partial v} 
- \frac{u^2}{\ell} \sin \theta \; e^{\pm i \phi} \: 
\frac{\partial}{\partial u} 
+ \frac{2 \: \ell \: u}{u v + \ell^2} \, \cos \theta \; e^{\pm i \phi} \: 
\frac{\partial}{\partial \theta} \: 
\pm \: i \, \frac{2 \: \ell\:  u}{u v + \ell^2} 
\: \frac{e^{\pm i \phi}}{\sin \theta} \; \frac{\partial}{\partial \phi} \; , 
\label{eqn:KillingDSppm} \\ 
\xi^{\mu}_{\scriptscriptstyle (\parallel\pm)} 
\! & \! \! \! = \! \! \! & \! 
\frac{v^2}{\ell} \sin \theta \; e^{\pm i \phi} \: 
\frac{\partial}{\partial v} - \ell \, \sin \theta \; e^{\pm i \phi} \: 
\frac{\partial}{\partial u} 
- \frac{2 \: \ell \: v}{u v + \ell^2} \, \cos \theta \; e^{\pm i \phi} \: 
\frac{\partial}{\partial \theta} \: 
\mp \: i \, \frac{2 \: \ell \: v}{u v + \ell^2} 
\: \frac{e^{\pm i \phi}}{\sin \theta} \; \frac{\partial}{\partial \phi} \; , 
\label{eqn:KillingDSmpm} \\ 
\xi^{\mu}_{\scriptscriptstyle (\pm)} 
\! & \! \! \! = \! \! \! & \! 
\pm e^{\pm i \phi} \: \frac{\partial}{\partial \theta} 
+ \: i \, \cot \theta \; e^{\pm i \phi} \: 
\frac{\partial}{\partial \phi} \; ,
\label{eqn:KillingDS2pm} \\ 
\xi^{\mu}_{\scriptscriptstyle (\phi)} 
\! & \! \! \! = \! \! \! & \! 
\frac{\partial}{\partial \phi} \; .  
\label{eqn:KillingDS34} 
\end{eqnarray}

\vskip 2cm
\baselineskip .2in



\begin{thebibliography}{99} 
\bibitem{Dbrane} A. Strominger and C. Vafa, 
Phys. Lett. {\bf B379}, 99 (1996); 
see, for review, G. T. Horowitz, preprint gr-qc/9604051. 
\bibitem{AdSCFT} J. Maldacena, 
Adv. Theor. Math. Phys. {\bf 2}, 231 (1998); 
S. S. Gubser, I.R. Klebanov and A. M. Polyakov, 
Phys. Lett. {\bf B428}, 105 (1998); 
E. Witten, Adv. Theor. Math. Phys. {\bf 2}, 253 (1998); 
O. Aharony, S. S. Gubser, J. Maldacena, H. Ooguri and Y. Oz, 
Phys. Rep. {\bf 323}, 183 (1998). 
\bibitem{TrappingHorizon} S. A. Hayward, 
Phys. Rev. D {\bf 49}, 6467 (1994); 
Class. Quantum Grav. {\bf 11}, 3025 (1994); 
Class. Quantum Grav. {\bf 15}, 3147 (1998). 
\bibitem{IsolatedHorizon} A. Ashtekar, C. Beetle and S. Fairhurst, 
Class. Quantum Grav. {\bf 17}, 253 (2000). 
\bibitem{WeaklyIsolated} A. Ashtekar, S. Fairhurst and B. Krishnan, 
Phys. Rev. D {\bf 62}, 104025 (2000); 
A. Ashtekar, C. Beetle and J. Lewandowski, preprint gr-qc/0103026. 
\bibitem{LQGentropy} A. Ashtekar, J. Baez, A. Corichi and K. Krasnov, 
Phys. Rev. Lett. {\bf 80}, 904 (1998). 
\bibitem{Strominger} A. Strominger, JHEP {\bf 02}, 009 (1998). 
\bibitem{BrownHenneaux} J. D. Brown and M. Henneaux, 
Comm. Math. Phys. {\bf 104}, 207 (1986). 
\bibitem{BMS}  H. Bondi, M. G. J. van der Burg and A. W. K. Metzner, 
Proc. Roy. Soc. (London) {\bf A269}, 21 (1962); 
R. K. Sachs, Proc. Roy. Soc. (London) {\bf A270}, 103 (1962); 
R. K. Sachs, Phys. Rev. {\bf 128}, 2851 (1962). 
\bibitem{Spi} A. Ashtekar and R. O. Hansen, 
J. Math. Phys. {\bf 19}, 1542 (1978); 
A. Ashtekar, in {\it General Relativity and Gravitation, Vol.2},  
ed. A. Held (New York, Plenum). 
\bibitem{Carlip1} S. Carlip, Phys. Rev. Lett. {\bf 82}, 2828 (1999). 
\bibitem{Carlip2} S. Carlip, 
Class. Quantum Grav. {\bf 16}, 3327 (1999). 
\bibitem{ParkHo} M-I. Park and J. Ho, 
Phys. Rev. Lett. {\bf 83}, 5595 (1999). 
\bibitem{CarpipRep} S. Carlip, Phys. Rev. Lett. {\bf 83}, 5596 (1999). 
\bibitem{DreyerEtAl} O. Dreyer, A. Ghosh and J. Wisniewski, preprint 
hep-th/0101117. 
\bibitem{GibbonsHawking} G. W. Gibbons and S. W. Hawking, 
Phys. Rev. D {\bf 15}, 2738 (1977). 
\bibitem{RaczWald} I. Racz and R. M. Wald, Class. Quantum Grav. 
{\bf 9} 2643 (1992). 
\bibitem{CovariantPhase} J. Lee and R. M. Wald, J. Math. Phys. 
{\bf 31}, 725 (1990); 
V. Iyer and R. M. Wald, Phys. Rev. D {\bf 50}, 846 (1994); 
V. Iyer and R. M. Wald, Phys. Rev. D {\bf 52}, 4430 (1995). 
\bibitem{WaldZoupas} R. M. Wald and A. Zoupas, Phys. Rev. D {\bf 61}, 
084027 (2000). 
\bibitem{HawkingRad} S. W. Hawking, Commun. Math. Phys. {\bf 43}, 
199 (1975). 
\bibitem{EuclideanBH} G. W. Gibbons and S. W. Hawking, Phys. Rev. D 
{\bf 15}, 2752 (1977); S. Hawking, in {\it General Relativity}, 
ed. S. Hawking and W. Israel (Cambridge University Press, Cambridge, 
1979). 
\bibitem{BirrellDavies} N. D. Birrell and P. C. W. Davies, {\it 
Quantum fields in curved space} (Cambridge University Press, Cambridge, 
1982). 
\bibitem{HottaEtAl} M. Hotta, K. Sasaki and T. Sasaki, 
preprint gr-qc/0011043, to appear in Class. Quantum. Grav.  
\bibitem{CarlipAdS} S. Carlip, Phys. Rev. D {\bf 51}, 
632 (1995). 
\bibitem{MaldacenaStrominger} J. Maldacena and A. Strominger, 
JHEP {\bf 02}, 014 (1998). 
\bibitem{Strominger2} A. Strominger, preprint hep-th/0106113. 
\bibitem{Solodukhin} S. N. Solodukhin, Phys. Lett. {\bf B454} 213 (1999). 
\bibitem{SachsSolo} I. Sachs and S. N. Solodukhin, 
preprint hep-th/0107173
\end{thebibliography}
\end{document}